\documentclass[aps,pra,preprint]{revtex4-1}
\usepackage{hyperref} 

\usepackage{graphicx} 
\usepackage{color}
\usepackage{filecontents}
\usepackage{soul}
\usepackage{physics}
\usepackage{mathrsfs}
\usepackage{subcaption}
\usepackage{relsize}
\usepackage{amsmath}
\usepackage{xcolor}
\usepackage{color, colortbl}
\definecolor{LightCyan}{rgb}{0.88,1,1}
\definecolor{corn}{rgb}{0.98, 0.93, 0.36}
\definecolor{pastelyellow}{rgb}{0.99, 0.99, 0.59}
\usepackage{tabularx,stackengine}

\usepackage{dcolumn}
\usepackage{bm}
\usepackage{amssymb,amsmath}
\usepackage{color}
\usepackage{float}
\usepackage{tikz}
\usetikzlibrary{arrows}

\begin{document}
\title{Two-photon superradiance and subradiance}
\author{Wenxuan Xie}
\affiliation{Department of Applied Physics, Yale University, New Haven, CT 06511, USA}
\email{wenxuan.xie@yale.edu}
\author{Imran M. Mirza}
\affiliation{Macklin Quantum Information Sciences, Department of Physics, Miami University, Oxford, Ohio 45056, USA}
\email{mirzaim@miamioh.edu}
\author{John C. Schotland}
\affiliation{Department of Mathematics and Department of Physics, Yale University, New Haven, CT 06511, USA} 
\email{john.schotland@yale.edu}

\date{\today}


\begin{abstract}
We consider the problem of two-photon cooperative emission in systems of two-level atoms. Two physically distinct regimes are analyzed. First, we investigate the case of a small number of atoms. We study the evolution of two-photon super- and sub-radiant states and associated two-photon spectra. Second, we investigate the problem of a constant density of atoms confined to a spherical volume. We analyze separately the cases in which the radius is small or large in comparison to the resonant wavelength. 
\end{abstract}

\maketitle

\section{\label{sec:1}Introduction}
Since the seminal work of Dicke~\cite{dicke1954coherence}, cooperative or collective spontaneous emission of electromagnetic radiation from atomic systems has remained a topic of fundamental interest and considerable practical importance. The key phenomenon, known as superradiance, occurs when a collection of $N$ atoms, whose size is smaller than the resonant wavelength of the electromagnetic radiation, emits light at an intensity  proportional to $N^2$ rather than $N$. Superradiant emission occurs on a timescale much shorter than the single-atom lifetime. This effect is due to the coupling of the atoms with the quantized field \cite{rehler1971superradiance, gross1982superradiance, scully2009super}. In contrast to superradiance, subradiance---the phenomenon that diminishes radiative emission--- leads to a much longer time scale for emission compared to the single-atom lifetime~\cite{crubellier1985superradiance}. Superradiance has been experimentally observed in a variety of physical systems, at wavelengths ranging from visible light to microwaves~\cite{skribanowitz1973observation,marek1979observation,gross1976observation,moi1983rydberg,crubellier1981experimental,nataf2010no,zeeb2015superradiant,chalony2011coherent,goban2015superradiance,roof2016observation,mlynek2014observation, das2020subradiance,gold2022spatial}. 

The recent availability of single-photon sources has led to the study of superradiance and subradiance at the single-photon level \cite{eisaman2011invited,rohlsberger2010collective, tighineanu2016single, de2014single, mirza2016fano}. To understand superradiance in this setting, we consider a system of $N$ atoms on which a single photon is incident. If the system is uniformly illuminated, information about which atom is excited is ultimately lost. It follows that the corresponding rate of spontaneous emission is increased by a factor of $N$ in comparison to that of a single-atom. This enhanced rate of single-photon emission has numerous applications in quantum information processing~\cite{an2009quantum, flamini2018photonic, northup2014quantum,solano2017super, lambert2016superradiance, norcia2018cavity,scully2015single,wang2020controllable,guimond2019subradiant}.

In this paper, we investigate the problem of two-photon cooperative emission. In contrast to the single-photon case, this problem is relatively unexplored, although we draw attention to the works~\cite{svidzinsky2010cooperative,Joe2}. We find that in the two-photon setting, the phenomena of superradiance and subradiance are extremely rich. The key idea is that the presence of a second photon leads to intrinsically quantum mechanical effects due to entanglement. These include photon-photon correlations~\cite{hong1987measurement}, photon-photon and photon-atom entanglement~\cite{mirza2016two}, and two-photon blockade \cite{hamsen2017two}. The scope of applications is also potentially enlarged and includes quantum communications, quantum networking, and optical imaging \cite{prabhakar2020two}. 

We now summarize our main results. We consider a collection of two level atoms interacting with a quantized field. Throughout this paper, we work within the rotating-wave and Wigner-Weisskopf approximations~\cite{scully1999quantum}. For simplicity, we restrict attention to a scalar model of the electromagnetic field and ignore the contributions of the Lamb shift. We begin with the single-atom case, thereby recovering the theory of stimulated emission. Next we analyze the two-atom case, where
we study the formation and evolution of two-photon super- and subradiant states in some detail. We also investigate the associated photon-photon correlations and spectral effects.
Finally, we consider a constant density of atoms in a spherical cavity. We analyze the cases when the cavity radius is small or large in comparison to the resonant wavelength $2\pi c / \Omega$, where $\Omega$ is the atomic resonance frequency. In both cases, we investigate the time-dependence of the emitted light and the two-photon spectrum. In particular,  we find that photon-photon correlations are stronger for smaller system sizes. In a related manner, we compute the von Neumann entropy as quantitative measure of two-photon entanglement. This provides an additional measure of correlations beyond the two-photon spectrum. Lastly, we find that the radiated power for two-photon emission is proportional to the number of atoms $N$ instead of $N^{2}$, as in Dicke superradiance, and has a nonlinear dependence on the system size. 

The paper is organized as follows. In Sec.~\ref{sec:model}, we describe the model system  and derive the equations of motion for the probability amplitudes that specify a two-photon state. In Sec.~\ref{sec:single} and Sec.~\ref{sec:two}, we study discrete atomic systems, especially for the case of two atoms. In Sec.~\ref{sec:small system} and Sec.~\ref{sec:large systems}, we investigate the problem of many atoms in a small cavity and large cavity, respectively. Finally, in Sec.~\ref{sec:discussion} we summarize our results and discuss future research directions. The appendices present the details of several calculations.


\section{\label{sec:model}Model}
We consider the following model for the interaction between a quantized field and a system of identical two-level atoms~\cite{dicke1954coherence}. The atoms, which are sometimes referred to as emitters, are assumed to be stationary and sufficiently well separated that interatomic interactions can be neglected. For simplicity, we adopt a scalar model of the electromagnetic field. The system is described by the Hamiltonian 
\begin{align}\label{HamTT}
\hat{H} =   \sum\limits_{\mathbf{k}} \hbar \omega_{\mathbf{k}} \hat{a}_{\mathbf{k}}^{\dagger} \hat{a}_{\mathbf{k}}  +  \sum\limits_j \hbar \Omega \hat{\sigma}^{\dagger}_j \hat{\sigma}_j + \hbar g \sum\limits_j \sum\limits_{\mathbf{k}} (e^{i \mathbf{k} \cdot \mathbf{r}_j}\hat{\sigma}^{\dagger}_j \hat{a}_{\mathbf{k}} + e^{-i \mathbf{k} \cdot \mathbf{r}_j} \hat{\sigma}_j \hat{a}_{\mathbf{k}}^{\dagger}) .
\end{align}
The Hamiltonian in Eq.~\eqref{HamTT} consists of three terms. The first term describes the Hamiltonian of the field, where $\omega_{\mathbf{k}}=c|\mathbf k|$ is the frequency of the field mode with wavevector $\mathbf{k}$ and $\hat{a}_{\mathbf{k}}^{\dagger}$ ($\hat{a}_{\mathbf{k}}$) is the corresponding creation (annihilation) operator. The operators $\hat{a}_{\mathbf{k}}^{\dagger}$ and $\hat{a}_{\mathbf{k}}$ obey the commutation relations for a bose field:
\begin{align}
\left[\hat{a}_{\mathbf{k}},\hat{a}_{\mathbf{p}}^{\dagger} \right] =\delta_{\mathbf{k} \mathbf{p}} \ , \quad
 \left[\hat{a}_{\mathbf{k}},\hat{a}_{\mathbf{p}}\right] = 0 .
\end{align}

The second term in Eq.~\eqref{HamTT} is the Hamiltonian of the atoms, where
$\Omega$ is the atomic transition frequency and $\hat{\sigma}^{\dagger}_j $ ($\hat{\sigma}_j$) is the atomic raising (lowering) operator of the $j$th atom. The operators $\hat{\sigma}^{\dagger}_j$ and $\hat{\sigma}_j$ obey the anticommutation relations
\begin{align}
    \big\{\hat{\sigma}_i,\hat{\sigma}_i^\dagger \big\} = 1 \ , \quad 
    \big\{\hat{\sigma}_i,\hat{\sigma}_i \big\} = 0  \ ,
    \label{commutator: on-site} 
\end{align}
along with the commutation relations
\begin{align}
    \big[\hat{\sigma}_i,\hat{\sigma}_j^\dagger \big] =0 \ , \quad \big[\hat{\sigma}_i,\hat{\sigma}_j\big] = 0 \ , \quad  i\neq j \ .  
     \label{commutator: off-site} 
\end{align}
That is, the atomic operators anticommute for the same atom and commute for different atoms. These mixed fermionic-bosonic commutation relations prohibit the double excitation of an atom while allowing the transfer of an excitation from one atom to another.
See Appendix~\ref{appd:Derivation of commutation relation for atomic operators} for further details.

The third term in Eq.~\eqref{HamTT} is the Hamiltonian that governs the interaction between the atoms and the field, where $\mathbf{r}_j$ is the position of the $j$th atom and $g$ is the atom-field coupling. We note that we have not included any counter rotating terms, consistent with the rotating wave approximation (RWA). In addition, $g$ is taken to be frequency-independent~\cite{scully1999quantum}. 
 
We suppose that the system is in a two-excitation state of the form
\begin{align}
\ket{\Psi(t)} = \left(  \sum\limits_{i,j} a_{ij}(t) \hat{\sigma}_{i}^{\dagger}\hat{\sigma}_{j}^{\dagger}+  \sum\limits_{i , \mathbf{k}} b_{i\mathbf{k}}(t) \hat{\sigma}_{i}^{\dagger} \hat{a}_{\mathbf{k}}^{\dagger} + \sum\limits_{\mathbf{k},\mathbf{p}} c_{\mathbf{k}\mathbf{p}}(t) \hat{a}_{\mathbf{k}}^{\dagger} \hat{a}_{\mathbf{p}}^{\dagger} \right) \lvert 0  \rangle,
    \label{eq: def of state}
\end{align}
where $\lvert 0 \rangle$ is the combined vacuum state of the field and the ground states of the atoms. Here $a_{ij}(t)$ is the probability amplitude of exciting atoms $i$ and $j$
at time $t$, $b_{i\mathbf{k}}(t)$ is the probability amplitude of exciting atom $i$ and creating a photon with wavevector $\mathbf{k}$ at time $t$, and $c_{\mathbf{k} \mathbf{p}}(t)$ is the probability amplitude of creating two photons with wave vectors $\mathbf{k}$ and $\mathbf{p}$ at time $t$. The following constraints on the probability amplitudes
\begin{align}
a_{ij}(t)  = a_{ji}(t), \quad a_{ii}(t) = 0, \quad c_{\mathbf{k} \mathbf{p}}(t) = c_{\mathbf{p}\mathbf{k}}(t)
\label{eq: constraint of the state}
\end{align}
follow from the commutation relations. Using the definition of the state and the above constraints, we define the following mode-independent probabilities:
\begin{equation}
    \left\vert a(t) \right\vert^{2} = 2 \sum\limits_{i,j} \left\vert a_{ij}(t) \right\vert^{2} , ~ \left\vert b(t) \right\vert^{2} =  \sum\limits_{i,\mathbf{k}} \left\vert b_{i \mathbf{k}}(t) \right\vert^{2}, ~  \left\vert  c(t) \right\vert^{2} =  2 \sum\limits_{\mathbf{k} , \mathbf{p}} \left\vert c_{\mathbf{k}\mathbf{p}}(t) \right\vert^{2} ,
    \label{eq:def of mode ind prob}
\end{equation}
in terms of which the conservation of probability is expressed as $\left\vert a(t) \right\vert^{2} + \left\vert b(t) \right\vert^{2} + \left\vert c(t) \right\vert^{2}   =1 $.

The dynamics of $\ket{\Psi(t)}$ is governed by the   equation 
\begin{align}
i\hbar\frac{\partial}{\partial t}\ket{\Psi}=\hat{H}\ket{\Psi} .
\end{align}
Projecting from the left-hand side by $\bra{0}\hat{\sigma}_i\hat{\sigma}_j$, $\bra{0}\hat{\sigma}_i\hat{a}_{\vb k}$ and $\bra{0}\hat{a}_{\vb k}\hat{a}_{\vb p}$ and making use of the atomic and field commutation relations, we find that the probability amplitudes obey the following system of equations:
\begin{subequations}
    \begin{align}
            i \frac{d }{dt}a_{ij}(t)  &= 2 \Omega a_{ij}(t) + \frac{g}{2}  \sum\limits_{\mathbf{k}}  b_{i \mathbf{k}}(t) e^{i \mathbf{k} \cdot \mathbf{r}_{j}} + \frac{g}{2} \sum\limits_{\mathbf{k}}  b_{j \mathbf{k}}(t) e^{i \mathbf{k} \cdot \mathbf{r}_{i}} - \delta_{ij} g \sum\limits_{\mathbf{k}} b_{i \mathbf{k}}(t) e^{i \mathbf{k} \cdot  \mathbf{r}_{i}},\\ 
            i \frac{d }{d t}b_{i \mathbf{k}}(t) &= \left(\Omega + \omega_{\mathbf{k}}\right) b_{i \mathbf{k}}(t) + 2 g \sum\limits_{j}  a_{ij}(t)e^{-i \mathbf{k} \cdot \mathbf{r}_{j}} +2g \sum\limits_{\mathbf{p}}   c_{\mathbf{k} \mathbf{p}}(t) e^{i\mathbf{p}\cdot\mathbf{r}_{i}},\\
            i \frac{d }{dt}c_{\mathbf{k} \mathbf{p}}(t) &= \left(\omega_{\mathbf{k}} + \omega_{\mathbf{p}}\right) c_{\mathbf{k} \mathbf{p}}(t) + \frac{g}{2} \sum\limits_{i}  b_{i \mathbf{k}}(t) e^{- i \mathbf{p} \cdot \mathbf{r}_{i}} + \frac{g}{2} \sum\limits_{i}  b_{i \mathbf{p}}(t) e^{-i \mathbf{k} \cdot \mathbf{r}_{i}}. 
\end{align}\label{eq:equation of motion} 
\end{subequations}
The derivation of the above equations is presented in Appendix \ref{appd:Equation of Motion}.


\section{\label{sec:single}Single atom problem}
In this section we consider the problem of a single atom, which serves to illustrate our results in the simplest setting. As may be expected, we recover the theory of stimulated emission, in which a photon interacts with an atom in its excited state~\cite{scully1999quantum}. Evidently, in this setting, Eqs.~\eqref{eq:equation of motion} become
\begin{subequations}
    \begin{align}
    i \frac{d  
     }{d t}b_{\mathbf{k}}(t) &=  (\Omega + \omega_{\mathbf{k}}) b_{\mathbf{k}}(t) + 2 g \sum\limits_{\mathbf{p}} c_{\mathbf{k}\mathbf{p}}(t), \\
    i \frac{d }{d t}c_{\mathbf{k}\mathbf{p}}(t) &=  (\omega_{\mathbf{k}}+\omega_{\mathbf{p}}) c_{\mathbf{k}\mathbf{p}}(t) + \frac{1}{2} g \left(b_{\mathbf{k}}(t)+b_{\mathbf{p}}(t)\right)\label{eq:equation of motion of single atom caseb},
\end{align}
\label{eq:equation of motion of single atom case}
\end{subequations}
where we have placed the atom at the origin, allowing us to omit the atomic index for simplicity. The conservation of probability is expressed as:
\begin{align}
\label{eq:conservation probability one atom}
  \sum\limits_{\mathbf{k}} \left\vert b_{\mathbf{k}}(t) \right\vert^{2}+2 \sum\limits_{\mathbf{k} , \mathbf{p}} \left\vert c_{\mathbf{k}\mathbf{p}}(t) \right\vert^{2}=1 .
\end{align}

We assume that the atom is initially excited and that there is a single photon with wavevector $\mathbf{k}_1$ in the field. This corresponds to the initial conditions $b_{\mathbf{k}}(0) = \delta_{\mathbf{k}\mathbf{k}_1}$ and $c_{\mathbf{k} \mathbf{p}}(0) = 0$. Eqs.~\eqref{eq:equation of motion of single atom case} can be solved by Laplace transforms. We find that 
\begin{subequations}
    \begin{align}
      &i \left( s b_{\mathbf{k}}(s) - b_{\mathbf{k}}(0)\right) =  (\Omega + \omega_{\mathbf{k}}) b_{\mathbf{k}}(s) + 2 g \sum\limits_{\mathbf{p}} c_{\mathbf{k}\mathbf{p}}(s), \\
    &i s c_{\mathbf{k}\mathbf{p}}(s) =  (\omega_{\mathbf{k}}+\omega_{\mathbf{p}}) c_{\mathbf{k}\mathbf{p}}(s) + \frac{1}{2} g \left(b_{\mathbf{k}}(s)+b_{\mathbf{p}}(s)\right).  
    \end{align}\label{eq:Laplace equation of motion of single atom case}
\end{subequations}
Here we have defined the Laplace transform by
\begin{equation}
    f(s) = \int_{0}^{\infty} e^{-s t} f(t) dt ,
\end{equation}
where $\Re(s) > 0$ and for convenience we denote a function and its Laplace transform by the same symbol. Next, we eliminate $c_{\mathbf{k}\mathbf{p}}(s)$ from Eq.~\eqref{eq:Laplace equation of motion of single atom case} and solve for $b_{\mathbf{k}}(s)$, which yields
\begin{align}
    b_{\mathbf{k}}(s) = \frac{i b_{\mathbf{k}}(0)  }{is -\Omega-\omega_{\mathbf{k}} -  \Sigma(s,\omega_\mathbf{k}) } + \frac{g^2}{is -\Omega-\omega_{\mathbf{k}} -  \Sigma(s,\omega_\mathbf{k})}\sum\limits_{\mathbf{p}}\frac{b_{\mathbf{p}}(s)}{is - \omega_{\mathbf{k}} - \omega_{\mathbf{p}}},
    \label{eq:self-consitent equation of single atom case}
\end{align}
where the self-energy $\Sigma(s,\omega)$ is defined by
\begin{align}
    \Sigma(s,\omega) = g^2 \sum\limits_{\mathbf{p}}\frac{1}{is -\omega -\omega_{\mathbf{p}}}
    \label{expression of self-energy}.
\end{align}
Inverting the Laplace transform in Eq.~\eqref{eq:self-consitent equation of single atom case}, we obtain
\begin{align}
\label{eq:inverse}
 b_{\mathbf{k}}(t) = \frac{1}{2\pi i } \int_C  e^{st}  b_{\mathbf{k}}(s) ds,
\end{align}
where the contour of integration $C$ is parallel to the imaginary axis in the complex $s$-plane, lying to the right of any singularities of the integrand (note that $\Sigma(s,\omega_{\mathbf{k}})$   decays as $1/s$ for large $|s|$).  In order to carry out the above integral, we make the pole approximation in which we replace $s$ with the pole $-i(\Omega + \omega_{\mathbf{k}})$ in $\Sigma(s,\omega_{\mathbf{k}})$ of Eq. \eqref{eq:self-consitent equation of single atom case}. This quantity is independent of $\mathbf k$ and thus we will denote it by $\Sigma$. We note that the pole approximation arises in the Wigner-Weisskopf theory of spontaneous emission~\cite{lambropoulos2007fundamentals}. We obtain that
\begin{align}
    b_{\mathbf{k}}(s) = \frac{i b_{\mathbf{k}}(0)  }{is -\Omega-\omega_{\mathbf{k}} - \Sigma} + \frac{g^2}{is -\Omega-\omega_{\mathbf{k}} - \Sigma}\sum\limits_{\mathbf{p}}\frac{  b_{\mathbf{p}}(s)}{is - \omega_{\mathbf{k}} - \omega_{\mathbf{p}}} .
\label{eq:self-consitent equation of single atom case approx}
\end{align}
We will find it useful to split $\Sigma$ into its real and imaginary parts according to
$\Sigma = \delta \omega - i{\Gamma}/{2}$. Here $\delta\omega$ is the Lamb shift and
\begin{equation}
\Gamma = \frac{g^2 V \Omega^2}{\pi c^3} ,
\label{def_Gamma}
\end{equation} 
where $V$ is the volume of the system. The quantity $\Gamma$ is the rate of spontaneous emission in scalar quantum electrodynamics~\cite{Joe1}.
A detailed calculation of $\Sigma$ is presented in Appendix~\ref{appd:calculation of self-energy}.

Eq.~\eqref{eq:self-consitent equation of single atom case approx} is a self-consistent equation for $b_{\mathbf{k}}(s)$, which we solve iteratively. We obtain to order $g^2$
that
\begin{align}
    b_{\mathbf{k}}(s) = \frac{i b_{\mathbf{k}}(0)}{is-\Omega-\omega_{\mathbf{k}}-\Sigma} +\frac{g^2}{is-\Omega-\omega_{\mathbf{k}} - \Sigma} \sum\limits_{\mathbf{p}} \frac{i 
    b_{\mathbf{p}}(0)}{(is-\omega_{\mathbf{k}}-\omega_{\mathbf{p}})(is-\Omega-\omega_{\mathbf{p}}-\Sigma)}+ \mathcal{O}(g^{4}).
  \label{general expression of b_vk of single-atom case}
\end{align}
We note that the above result holds for weak coupling with $g / \Omega \ll 1$. 
Imposing the initial condition $b_{\mathbf{k}}(0) = \delta_{\mathbf{k} \mathbf{k}_{1}}$, we observe that the summation over the momentum $\mathbf{p}$ in Eq.~\ref{general expression of b_vk of single-atom case} can be performed. We thus obtain
\begin{equation}
    b_{\mathbf{k}}(s) = \frac{i \delta_{\mathbf{k} \mathbf{k}_{1}}}{is-\Omega-\omega_{\mathbf{k}}-\Sigma} + \frac{i g^2}{(is - \Omega - \omega_{\mathbf{k}} -\Sigma)(is - \omega_{\mathbf{k}} - \omega_{\mathbf{k}_{1}})(is - \Omega - \omega_{\mathbf{k}_{1}} - \Sigma)}+\mathcal{O}(g^{4}).
\end{equation}
Finally, inverting the Laplace transform yields the required expression for $b_{\mathbf{k}}(t)$:
\begin{equation}
\begin{split}
    b_{\mathbf{k}}(t) = &\delta_{\mathbf{k} \mathbf{k}_{1}} e^{-i(\Omega  + \omega_{\mathbf{k}})t -\frac{\Gamma}{2}t} + g^2 \left[ \frac{ e^{-i(\omega_{\mathbf{k}}+\Omega )t -\frac{\Gamma}{2}t} }{ \left(\omega_{\mathbf{k}}-\omega_{\mathbf{k}_{1}} \right)  \left(\Omega -\omega_{\mathbf{k}_{1}}-i \Gamma/2  \right) }  + \frac{ e^{-i(\omega_{\mathbf{k}_{1}}+\Omega )t -\frac{\Gamma}{2}t} }{ \left(\omega_{\mathbf{k}_{1}}-\omega_{\mathbf{k}}\right)  \left(\Omega -\omega_{\mathbf{k}}-i \Gamma/2  \right) } \right. \\
    &\left. + \frac{ e^{-i(\omega_{\mathbf{k}}+\omega_{\mathbf{k}_{1}})t} }{ \left( \Omega -\omega_{\mathbf{k}} - i \Gamma/2  \right)  \left(\Omega -\omega_{\mathbf{k}_{1}}-i \Gamma/2  \right) } \right]+\mathcal{O}(g^{4}),
    \end{split}
    \label{eq:stimulated emission solution to one photon amplitude}
\end{equation}
where we have ignored the Lamb shift by absorbing it into the the transition frequency $\Omega$. Using  Eq. \eqref{eq:stimulated emission solution to one photon amplitude}, we obtain the two-photon amplitude $c_{\mathbf{k} \mathbf{p}}$ by integrating
Eq.~\eqref{eq:equation of motion of single atom caseb} over $t$:
\begin{align}
c_{\mathbf{k} \mathbf{p}}(t) = \frac{1}{2}g \delta_{\mathbf{k} \mathbf{k}_{1}}\frac{ e^{-i(\omega_{\mathbf{k}} + \omega_{\mathbf{p}}) t } - e^{-i(\Omega  + \omega_{\mathbf{k}})t -\frac{\Gamma}{2}t}}{\omega_{\mathbf{p}}-\Omega  + i \Gamma/2} +(\mathbf{k} \leftrightarrow \mathbf{p})+\mathcal{O}(g^{3}).
\label{eq:stimulated emission solution to two photon amplitude}
\end{align}
Follow the definition of mode-independent probabilities in Eq. \eqref{eq:def of mode ind prob}, we find
\begin{subequations}
    \begin{align}
   &\left\vert b(t) \right\vert^{2} = e^{- \Gamma  t}+ \frac{g^{2} e^{- \frac{1}{2} \Gamma t }}{\left[ (\Omega - \omega_{\mathbf{k}_{1}})^{2} + \frac{1}{4}\Gamma^{2} \right]^2} \left( (\Omega-\omega_{\mathbf{k}_{1}} + \frac{1}{2} i \Gamma)^{2} e^{i(\Omega - \omega_{\mathbf{k}_{1}})t} + \text{c.c.} \right) ,\\
   &\left\vert c(t) \right\vert^{2}  
    = 1- e^{-\Gamma t} +g^{2} \frac{1- 2e^{- \frac{1}{2} \Gamma t } \cos(\Omega - \omega_{\mathbf{k}_{1}})t + e^{- \Gamma t} }{(\Omega-\omega_{\mathbf{k}_{1}})^{2} + \frac{1}{4} \Gamma^{2}}.
    \end{align}
    \label{eq:stimulated emission probabilities}
\end{subequations}
\begin{figure}
    \centering
    \includegraphics[width = 0.7\textwidth]{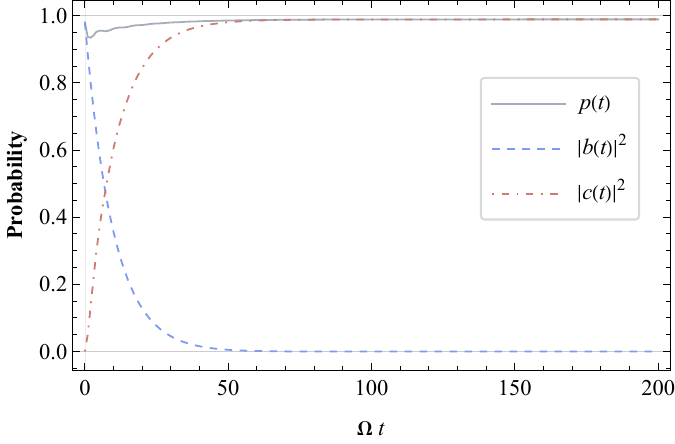}
    \captionsetup{
  format=plain,
  margin=1em,
  justification=raggedright,
  singlelinecheck=false
}
    \caption{Time evolution of the probabilities $ \left\vert b(t) \right\vert^{2} $, $ \left\vert c(t) \right\vert^{2} $ and the total probability $p(t)$. Here we have set $g / \Omega  =0.005$,  $\Gamma / \Omega  = 0.1$ and $\omega_{\mathbf{k}_{1}}  / \Omega = 1.0$.}
\label{fig:Single_Atom_all Probability}
\end{figure} 
In order to obtain Eq. \eqref{eq:stimulated emission probabilities}, the summation over modes has been replaced by an integral according to $\sum_{\mathbf{k}} \to V /(2\pi)^{3} \int d^{3}k$. 
We note that the resulting integral is divergent. To address this problem, we approximate the photonic density of states as being localized around the atomic resonant frequency. This allows us to evaluate the integral on-shell by replacing $\int d^{3} {k}$ by $k_0^2 \int dk  d \hat{\mathbf{k}}$, where $k_0 = \Omega / c$ is the wavenumber corresponding to the atomic transition frequency. Physically, this means that the atom interacts only with photons whose frequency is close to the atomic resonance frequency. 

In Fig.~\ref{fig:Single_Atom_all Probability}, we plot the quantities $|b(t)|^2$, $|c(t)|^2$ and the total probability $p(t) = |b(t)|^2 + |c(t)|^2$, as defined by Eq.~\eqref{eq:conservation probability one atom}. We observe that $|b(t)|^2$ and $|c(t)|^2$ decay and increase in time, respectively. We also note that $p(t)$ is not conserved at all times.
\begin{figure}[t]
\centering
    \subfloat[\label{fig:Stimulated_Emission_Two_Photon_Spectrum}]{\includegraphics[width =0.4\textwidth]{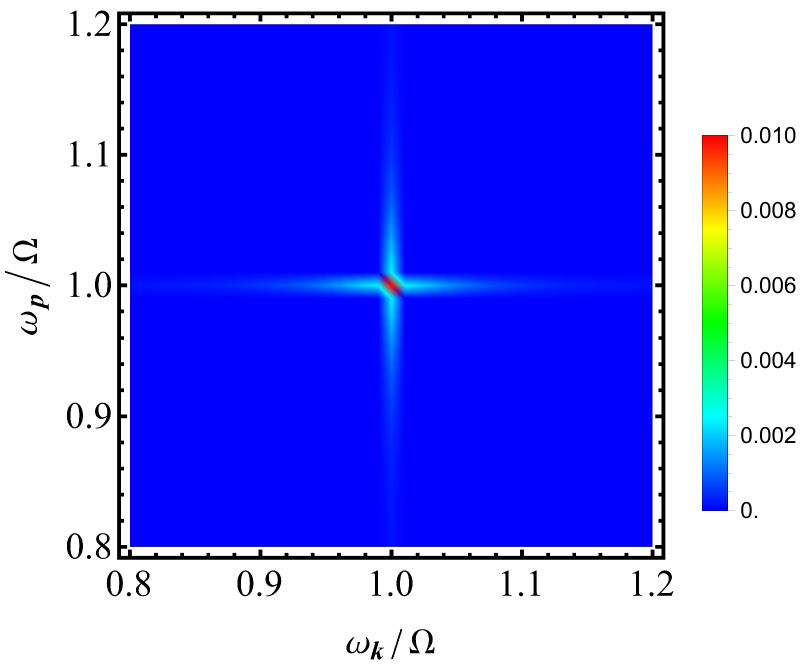}}
    \subfloat[\label{fig:Stimulated_Emission_Power}]{\includegraphics[width =0.5\textwidth]{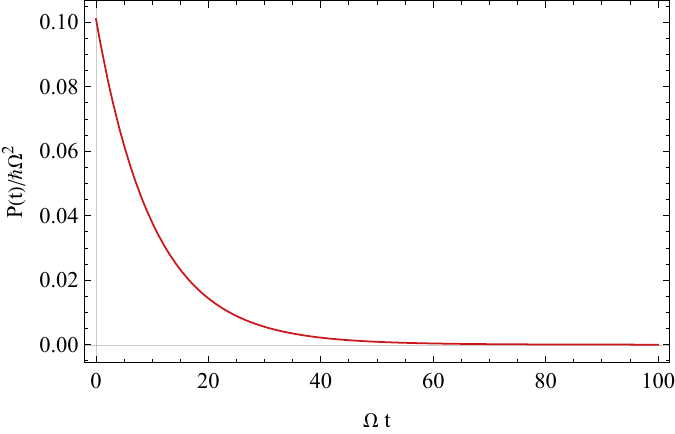}}
    
    \captionsetup{
  format=plain,
  margin=1em,
  justification=raggedright,
  singlelinecheck=false
}
    \caption{(a)Two photon spectrum (b) Radiated power for the single-atom problem. All parameters are the same as Fig. \ref{fig:Single_Atom_all Probability}.}
\end{figure}

To study the two-photon emission spectrum, we take $t\to\infty$ and obtain the following limiting behavior of Eq.~\eqref{eq:stimulated emission solution to two photon amplitude}:
\begin{equation}
     c_{\mathbf{k}\mathbf{p}}(t \to \infty) = \frac{1}{2}g e^{-i(\omega_{\mathbf{k}}+\omega_{\mathbf{p}})t} \left( \frac{\delta_{\mathbf{k}\mathbf{k}_{1}}}{\omega_{\mathbf{p}}-\Omega+i\frac{\Gamma}{2}} +\frac{\delta_{\mathbf{p}\mathbf{k}_{1}}}{\omega_{\mathbf{k}}-\Omega+i\frac{\Gamma}{2}}\right)+\mathcal{O}(g^{3}).
\end{equation}
In Fig. \ref{fig:Stimulated_Emission_Two_Photon_Spectrum} we plot the two-photon spectrum $\rho(\omega_{\mathbf{k}},\omega_{\mathbf{p}})= \left\vert c_{\mathbf{k}\mathbf{p}}(t \to \infty) \right\vert^{2} $ as a function of the photon frequencies $\omega_{\vb k}$ and $\omega_{\vb p}$. As expected, we find that the photons are not correlated. 

Next, we study the radiated power. To proceed, we calculate the time dependence of photon energy in the field, which is defined as 
\begin{equation}
  E(t) = \sum\limits_{\mathbf{k}} \hbar \omega_{\mathbf{k}} \left\vert b_{\mathbf{k}}(t) \right\vert^{2}  +2 \sum\limits_{\mathbf{k},\mathbf{p}} \hbar (\omega_{\mathbf{k}} +\omega_{\mathbf{p}})\left\vert c_{\mathbf{k}\mathbf{p}}(t) \right\vert^{2}.
    \label{eq:def of energy of single atom}
\end{equation}
Next we substitute Eqs. \eqref{eq:stimulated emission solution to one photon amplitude} and \eqref{eq:stimulated emission solution to two photon amplitude} into Eq. \eqref{eq:def of energy of single atom}, and convert the summation into an integral. The integral over $\omega_{\mathbf{k}}$ is regularized by replacing the photon frequency $\omega_{\mathbf{k}}$ by the atom transition frequency  $\Omega$ by the on-shell approximation. We thus obtain
\begin{align}
    E(t)  = & \ \hbar(\Omega + \omega_{\mathbf{k}_{1}}) -\hbar \Omega e^{-\Gamma t} + \frac{2\hbar g^{2} \omega_{\mathbf{k}_{1}}}{(\Omega-\omega_{\mathbf{k}_{1}})^2 + \Gamma^2/4} \left[ 1 -\frac{\Gamma(\Omega-\omega_{\mathbf{k}_{1}}) e^{-\Gamma t /2} \sin{(\Omega -\omega_{\mathbf{k}_{1}})t} }{(\Omega-\omega_{\mathbf{k}_{1}})^2 + \Gamma^2/4}  \right. \nonumber \\
    & \left. - \left(1 + \frac{1}{2} \frac{\Gamma^2}{(\Omega-\omega_{\mathbf{k}_{1}})^2 + \Gamma^2/4} \right)e^{-\Gamma t /2} \cos{(\Omega -\omega_{\mathbf{k}_{1}})t} + 2 e^{-\Gamma t} \right].
\end{align}
The radiated power is defined as $P(t) = d E(t) / dt$, which is given by 
\begin{align}
    P(t) = &\hbar \Omega \Gamma e^{-\Gamma t} +  \frac{\hbar g^{2} \omega_{\mathbf{k}_{1}}}{(\Omega-\omega_{\mathbf{k}_{1}})^2 + \Gamma^2/4} \left[ \left( -1 + \frac{\Gamma^2}{(\Omega-\omega_{\mathbf{k}_{1}})^2 + \Gamma^2/4} \right) \Gamma e^{-\Gamma t /2} \cos{(\Omega -\omega_{\mathbf{k}_{1}}) t} \right. \nonumber \\
    &+ \left.2\left( 1 + \frac{\Gamma^2}{(\Omega-\omega_{\mathbf{k}_{1}})^2 + \Gamma^2/4} \right) (\Omega -\omega_{\mathbf{k}_{1}}) e^{-\Gamma t /2} \sin{(\Omega-\omega_{\mathbf{k}_{1}})t} - 2 \Gamma e^{-\Gamma t} \right].
\end{align}
The radiated power is plotted in Fig. \ref{fig:Stimulated_Emission_Power}. It can be seen that the power achieves its maximum value at $t=0$ and decays monotonically at long times.

\section{\label{sec:two}Two Atom Problem}
We now turn our attention to the case of two atoms. The equations of motion for the probability amplitudes follow from Eq.~\eqref{eq:equation of motion} and are of the form
\allowdisplaybreaks
\begin{subequations}
    \begin{align}
        i \frac{d }{dt}a_{12}(t) &= 2 \Omega a_{12}(t) + \frac{g}{2} \sum\limits_{\mathbf{k}} (b_{1\mathbf{k}}(t) e^{i \mathbf{k} \cdot \mathbf{r}_2}+ b_{2\mathbf{k}}(t) e^{i \mathbf{k} \cdot \mathbf{r}_1}), \\
    i \frac{ d }{dt}b_{1\mathbf{k}}(t) &= (\Omega+\omega_{\mathbf{k}})b_{1\mathbf{k}}(t) + 2 g e^{-i\mathbf{k} \cdot \mathbf{r}_2} a_{12}(t) + 2 g \sum\limits_{\mathbf{p}} e^{i\mathbf{p} \cdot \mathbf{r}_1} c_{\mathbf{k} \mathbf{p}}(t),\\
    i \frac{ d }{dt}b_{2\mathbf{k}}(t) &= (\Omega+\omega_{\mathbf{k}})b_{2\mathbf{k}}(t) + 2 g e^{-i\mathbf{k} \cdot \mathbf{r}_1} a_{21}(t) + 2 g \sum\limits_{\mathbf{p}} e^{i\mathbf{p} \cdot \mathbf{r}_2} c_{\mathbf{k} \mathbf{p}}(t),\\
    i \frac{d }{dt} c_{\mathbf{k} \mathbf{p}}(t) &= (\omega_{\mathbf{k}} + \omega_{\mathbf{p}}) c_{\mathbf{k} \mathbf{p}}(t) + \frac{g}{2} \left(b_{1\mathbf{k}}(t)e^{-i\mathbf{p} \cdot \mathbf{r}_1}+ b_{2\mathbf{k}}(t)e^{-i\mathbf{p} \cdot \mathbf{r}_2} + b_{1\mathbf{p}}(t)e^{-i\mathbf{k} \cdot \mathbf{r}_1} + b_{2\mathbf{p}}(t)e^{-i\mathbf{k} \cdot \mathbf{r}_2} \right).
    \end{align}
\label{eq:equation of motion two atoms}
\end{subequations}

We note that due to the symmetry condition $a_{12}(t) = a_{21}(t)$, the equation of motion for $a_{21}$ is redundant. We assume that both atoms are excited and that there are no photons present in the field. This corresponds to the initial conditions 
$a_{12}(0) = {1}/{2}, b_{1\mathbf{k}}(0)= b_{2\mathbf{k}}(0) = c_{\mathbf{k} \mathbf{p}}(0)= 0$. Note that $a_{12}(0)+a_{21}(0) = 1$, so that the state is properly normalized. We solve  Eq.~\eqref{eq:equation of motion two atoms} using the same technique and approximations as in the one-atom case. We begin by Laplace transforming Eq.~\eqref{eq:equation of motion two atoms} and applying the initial conditions. We thus
obtain
\begin{subequations}
    \begin{align}
    &i \left(s a_{12}(s) - \frac{1}{2}\right) = 2 \Omega a_{12}(s) + \frac{g}{2} \sum\limits_{\mathbf{k}} (b_{1\mathbf{k}}(s) e^{i \mathbf{k} \cdot \mathbf{r}_2}+ b_{2\mathbf{k}}(s) e^{i \mathbf{k} \cdot \mathbf{r}_1}), \label{equation of motion after Laplace transformation for two-atom casea} \\
    &i s b_{1\mathbf{k}}(s) = (\Omega+\omega_{\mathbf{k}})b_{1\mathbf{k}} (s) + 2 g e^{-i\mathbf{k} \cdot \mathbf{r}_2} a_{12}(s) + 2 g \sum\limits_{\mathbf{p}} e^{i\mathbf{p} \cdot \mathbf{r}_1} c_{\mathbf{k} \mathbf{p}}(s), \label{equation of motion after Laplace transformation for two-atom caseb1}\\
    &i s b_{2\mathbf{k}}(s) = (\Omega+\omega_{\mathbf{k}})b_{2\mathbf{k}}(s) + 2 g e^{-i\mathbf{k} \cdot \mathbf{r}_1} a_{12}(s) + 2 g \sum\limits_{\mathbf{p}} e^{i\mathbf{p} \cdot \mathbf{r}_2} c_{\mathbf{k} \mathbf{p}}(s), \label{equation of motion after Laplace transformation for two-atom caseb2}\\
    &i s c_{\mathbf{k} \mathbf{p}}(s) = (\omega_{\mathbf{k}} + \omega_{\mathbf{p}}) c_{\mathbf{k} \mathbf{p}}(s) + \frac{g}{2} \left(b_{1\mathbf{k}}(s) e^{-i\mathbf{p} \cdot \mathbf{r}_1}+ b_{2\mathbf{k}}(s) e^{-i\mathbf{p} \cdot \mathbf{r}_2} + b_{1\mathbf{p}}(s) e^{-i\mathbf{k} \cdot \mathbf{r}_1} + b_{2\mathbf{p}}(s) e^{-i\mathbf{k} \cdot \mathbf{r}_2} \right).
    \label{equation of motion after Laplace transformation for two-atom casec}
    \end{align}
\end{subequations}

To make further progress, we eliminate the amplitudes $b_{1\mathbf{k}}$ and $b_{2\mathbf{p}}$ that appear in Eq.~\eqref{equation of motion after Laplace transformation for two-atom casea}. We find that
\begin{equation}
    \left( is-2\Omega-2\Sigma\left(s,\Omega \right) \right)a_{12}(s) = \frac{i}{2} +g^{2} \sum\limits_{\mathbf{k},\mathbf{p}} \frac{e^{i(\mathbf{k}\cdot \mathbf{r}_{1}+\mathbf{p}\cdot \mathbf{r}_{2})} c_{\mathbf{k}\mathbf{p}}(s)}{is-\Omega-\omega_{\mathbf{k}}}+\left( 1 \leftrightarrow 2 \right).
\end{equation}
In the weak coupling regime ($g / \Omega \ll 1$), it follows from the above result that $a_{12}(s)$ is given by
\begin{equation}
    a_{12}(s) = \frac{1}{2}\frac{1}{s +2 i \Omega +2 i  \Sigma(s,\Omega)}+\mathcal{O}(g^4),
\end{equation}
since $b_{1\mathbf k}(s)$ and $b_{2\mathbf k}(s)$ are $\mathcal{O}(g)$ and $c_{\mathbf {kp}}(s)$ is $\mathcal{O}(g^2)$. We then make the pole approximation by replacing $s$ with the pole $-i 2\Omega$ in $\Sigma(s,\Omega)$. The corresponding quantity is denoted by $\Sigma = \delta \omega - i{\Gamma}/2$. The self-energy $\Sigma$ here is the same as the one in the one atom case, and the definition of $\Gamma$ is the same as Eq. \eqref{def_Gamma}. We obtain
\begin{equation}
    a_{12}(s) = \frac{1}{2}\frac{1}{s +2 i \Omega +2 i  \Sigma},
    \label{eq:two atom amplitude}
\end{equation}

Furthermore, by eliminating the amplitude $c_{\mathbf{kp}}$ in Eqs.~\eqref{equation of motion after Laplace transformation for two-atom caseb1} and \eqref{equation of motion after Laplace transformation for two-atom caseb2}, 
we see that to leading order in $g$, $b_{1\mathbf{k}}(s)$ and  $b_{2\mathbf{k}}(s)$ are of the form
\begin{subequations}
    \begin{align}
    (is-\Omega-\omega_{\mathbf{k}}- \Sigma(s,\omega_{\mathbf{k}} )  )   b_{1\mathbf{k}}(s) - \Delta(s,\omega_\mathbf{k}) b_{2\mathbf{k}}(s) &=  2  g e^{-i\mathbf{k} \cdot \mathbf{r}_2} a_{12}(s) + \mathcal{O}(g^{3}), \\
    (is-\Omega-\omega_{\mathbf{k}}- \Sigma(s,\omega_{\mathbf{k}} )  ) b_{2\mathbf{k}}(s) - \Delta(s,\omega_\mathbf{k}) b_{1\mathbf{k}}(s) &=  2 ge^{-i\mathbf{k} \cdot \mathbf{r}_1} a_{12}(s) + \mathcal{O}(g^{3}) ,
    \end{align}
    \label{eq:two atom mixed amplituded result}
\end{subequations}
where we have introduced the interaction energy $\Delta(s,\omega_\mathbf{k})$, which is defined by
\begin{equation}
        \Delta(s,\omega_\mathbf{k}) = g^2 \sum\limits_{\mathbf{p}}\frac{e^{i\mathbf{p} \cdot (\mathbf{r}_1 - \mathbf{r}_2)}}{is -\omega_{\mathbf{k}}-\omega_{\mathbf{p}}}.
        \label{def_Delta}
\end{equation}

Similarly, we will make the pole approximation in Eq. \eqref{eq:two atom mixed amplituded result}. But since the position of pole is changed to $\Omega+\omega_{\mathbf{k}}$ for amplitude $b_{1\mathbf{k}}(s)$, we replace $s$ with $-i(\Omega+\omega_{\mathbf{k}})$ in $\Sigma(s,\omega_{\mathbf{k}} ) $ and $\Delta(s,\omega_{\bf k})$, which gives us $\Sigma = \delta \omega - i{\Gamma}/2$ and $\Delta =\delta\omega_{r} - i {\Gamma} \textrm{sinc}{(k_{0}r)/2}$ respectively. The definition of $\Gamma$ is the same as that of Eq. \eqref{def_Gamma}, $r = \left\vert \mathbf{r}_{1} - \mathbf{r}_{2} \right\vert$ and $k_{0} = \Omega / c$. The definition of $\delta\omega_r$ and the calculation details of $\Delta$ are presented in Appendix \ref{appd:calculation of interaction energy}. We find that to leading order in $g$,  $b_{1\mathbf{k}}(s)$ and $b_{2\mathbf{k}}(s)$ become
\begin{subequations}
    \begin{align}
    b_{1\mathbf{k}}(s) &= g \frac{1}{s +2 i \Omega +2 i \Sigma} \left( \frac{e^{-i \mathbf{k} \cdot \mathbf{r}_1}+e^{-i\mathbf{k} \cdot \mathbf{r}_2}}{is-\Omega-\omega_{\mathbf{k}}-\Sigma-\Delta}-\frac{e^{-i\mathbf{k} \cdot \mathbf{r}_1}-e^{-i \mathbf{k} \cdot \mathbf{r}_2}}{is-\Omega-\omega_{\mathbf{k}}-\Sigma+\Delta} \right),\\
    b_{2\mathbf{k}}(s) &= g \frac{1}{s +2 i \Omega +2 i \Sigma} \left( \frac{e^{-i \mathbf{k} \cdot \mathbf{r}_1}+e^{-i\mathbf{k} \cdot \mathbf{r}_2}}{is-\Omega-\omega_{\mathbf{k}}-\Sigma-\Delta}+\frac{e^{-i\mathbf{k} \cdot \mathbf{r}_1}-e^{-i \mathbf{k} \cdot \mathbf{r}_2}}{is-\Omega-\omega_{\mathbf{k}}-\Sigma+\Delta} \right).
    \end{align}
\end{subequations}
Substituting the above into Eq.~\eqref{equation of motion after Laplace transformation for two-atom casec}, we obtain the following expression for $c_{\mathbf{k}\mathbf{p}}(s)$:
\begin{align}
    c_{\mathbf{k} \mathbf{p}}(s) &= \frac{  i g^2 }{4(is-\omega_{\mathbf{k}}-\omega_{\mathbf{p}})  
     } \left\{
    \frac{e^{-i(\mathbf{p} \cdot \mathbf{r}_1 + \mathbf{k} \cdot \mathbf{r}_2)} + e^{-i(\mathbf{k} \cdot \mathbf{r}_1 + \mathbf{p} \cdot \mathbf{r}_2)} + e^{-i(\mathbf{k}+\mathbf{p}) \cdot \mathbf{r}_1} + e^{-i(\mathbf{k}+\mathbf{p}) \cdot \mathbf{r}_2}}{(is-2\Omega-2\Sigma)(is-\omega_{\mathbf{k}} - \Omega - \Sigma - \Delta )}\right. \nonumber \\
    & \left. +\frac{e^{-i(\mathbf{p} \cdot \mathbf{r}_1 + \mathbf{k} \cdot \mathbf{r}_2)} + e^{-i(\mathbf{k} \cdot \mathbf{r}_1 + \mathbf{p} \cdot \mathbf{r}_2)} - e^{-i(\mathbf{k}+\mathbf{p}) \cdot \mathbf{r}_1} - e^{-i(\mathbf{k}+\mathbf{p}) \cdot \mathbf{r}_2}}{(is-2\Omega-2\Sigma)(is-\omega_{\mathbf{k}} - \Omega - \Sigma + \Delta )} \right\} + (\mathbf{k} \leftrightarrow \mathbf{p})+\mathcal{O}(g^4).
    \label{eq:amp ckp}
\end{align}
\begin{figure}
    \centering
    \includegraphics[width = 0.75\textwidth]{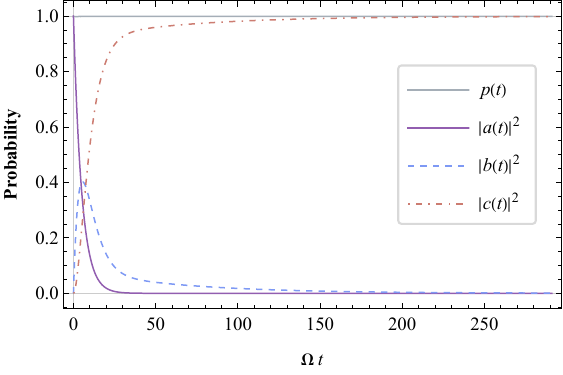}
    \captionsetup{
  format=plain,
  margin=1em,
  justification=raggedright,
  singlelinecheck=false
}
\caption{Time dependence of the probabilities $|a(t)|^2$, $|b(t)|^2, |c(t)|^2$ and $p(t)$. We set $\Gamma/\Omega =0.1$, $k_{0} \left| \mathbf{r}_1-\mathbf{r}_2 \right|=1.0$.}
    \label{fig:Two_Photon_all_probability_plot}
\end{figure}
Finally, performing the inverse Laplace transform and integrating over the modes,  we obtain the following expressions for the mode-independent probabilities:
\begin{subequations}
    \begin{align}
        &\left\vert a(t) \right\vert^{2} = e^{-2\Gamma t},\\
        &\left\vert b(t) \right\vert^{2} =  \frac{\Gamma_{+}}{\Gamma_{-}} (e^{-\Gamma_{+}t} - e^{-2\Gamma t})+\frac{\Gamma_{-}}{\Gamma_{+}} (e^{-\Gamma_{-}t} - e^{-2\Gamma t}) , \\
        &\left\vert c(t) \right\vert^{2} = 1 - \left(1 - \frac{\Gamma_{+}^{2} + \Gamma_{-}^{2}}{\Gamma_{+}\Gamma_{-}}\right) e^{-2\Gamma t } - \frac{\Gamma_{+}}{\Gamma_{-}} e^{-\Gamma_{+}t} -\frac{\Gamma_{-}}{\Gamma_{+}} e^{-\Gamma_{-}t} ,
    \end{align}\label{twoatomres}
\end{subequations}
where $\Gamma_{\pm} = \Gamma (1\pm \text{sinc}(k_0 r))$, with $r= \left| \mathbf{r}_1-\mathbf{r}_2 \right|$ the distance between the atoms. 
In obtaining the above result, the sum over modes has been converted to an integral, as was done in deriving Eq. \ref{eq:stimulated emission probabilities} and the Lamb shift has been ignored. The details of the calculations are presented in Appendix \ref{appd:two atom problem}.

In Fig.~\ref{fig:Two_Photon_all_probability_plot}, we present the time evolution of the probabilities $|a(t)|^2$, $|b(t)|^2$, and $|c(t)|^2$, along with the total probability $p(t)=|a(t)|^2 + |b(t)|^2 + |c(t)|^2$. The two-atom excitation probability $|a(t)|^2$ (solid purple curve) exhibits an exponential decay over time. In contrast, the single-photon mixed-state probability $|b(t)|^2$ (blue dashed curve) initially increases to a peak value before gradually decreasing. At long times, both $|a(t)|^2$ and $|b(t)|^2$ approach zero, while the two-photon probability $|c(t)|^2$ (red dotted-dashed curve) asymptotically approaches unity. Note that the total probability remains conserved throughout the evolution, which differs from the single-atom case. This may be attributed to the fact that in this setting, the system emits photons whose energies precisely match the atomic resonance frequency. Consequently, the replacement $\int d^{3} k \to k_{0}^{2} \int d k d \hat{\mathbf{k}}$ does not incur an error.

\subsection{Two-Photon Spectrum}
\label{spectrum two atom}
In this part, we will study the spectrum of two photons at long times. Since the atomic probability $\left\vert a(t) \right\vert^{2} $ and mixed state probability $\left\vert b(t) \right\vert^{2} $ vanish in the limit $t\to\infty$ , the final state of the system is a two-photon state.
Hence, by the Eq. \eqref{eq:amp ckp}, the total probability is of the form
\begin{equation}
   p(t\to\infty) = \left\vert c(t \to  \infty) \right\vert^{2} = \int_{-\infty}^{\infty} \int_{-\infty}^{\infty} d \omega_{\mathbf{k}} d \omega_{\mathbf{p}} \rho(\omega_{\mathbf{k}},\omega_{\mathbf{p}}),
    \label{eq:spectral_integral}
\end{equation}
where the two-photon spectral density $\rho(\omega_{\mathbf{k}},\omega_{\mathbf{p}})$ is defined by
\begin{equation}
       \rho(\omega_{\mathbf{k}},\omega_{\mathbf{p}}) =  \frac{\Gamma_{+}^{2} \left( \left( 2\Omega - \omega_{\mathbf{k}}-  \omega_{\mathbf{p}}\right)^{2}+\Gamma_{+}^{2} \right) / (8 \pi^{2})}{ \left( \left(2\Omega - \omega_{\mathbf{k}} - \omega_{\mathbf{p}}\right)^{2}  +\Gamma^2  \right) \left( (\Omega-\omega_{\mathbf{k}})^{2} + \frac{\Gamma_{+}^{2}}{4}\right) \left( (\Omega-\omega_{\mathbf{p}})^{2} +  \frac{\Gamma_{+}^{2}}{4} \right) } + ( \Gamma_{+} \leftrightarrow \Gamma_{-}).
    \label{eq: photon joint probability of two-atom case}
\end{equation} 

\begin{figure}[t]
    \centering
    \includegraphics[width = 0.9\textwidth]{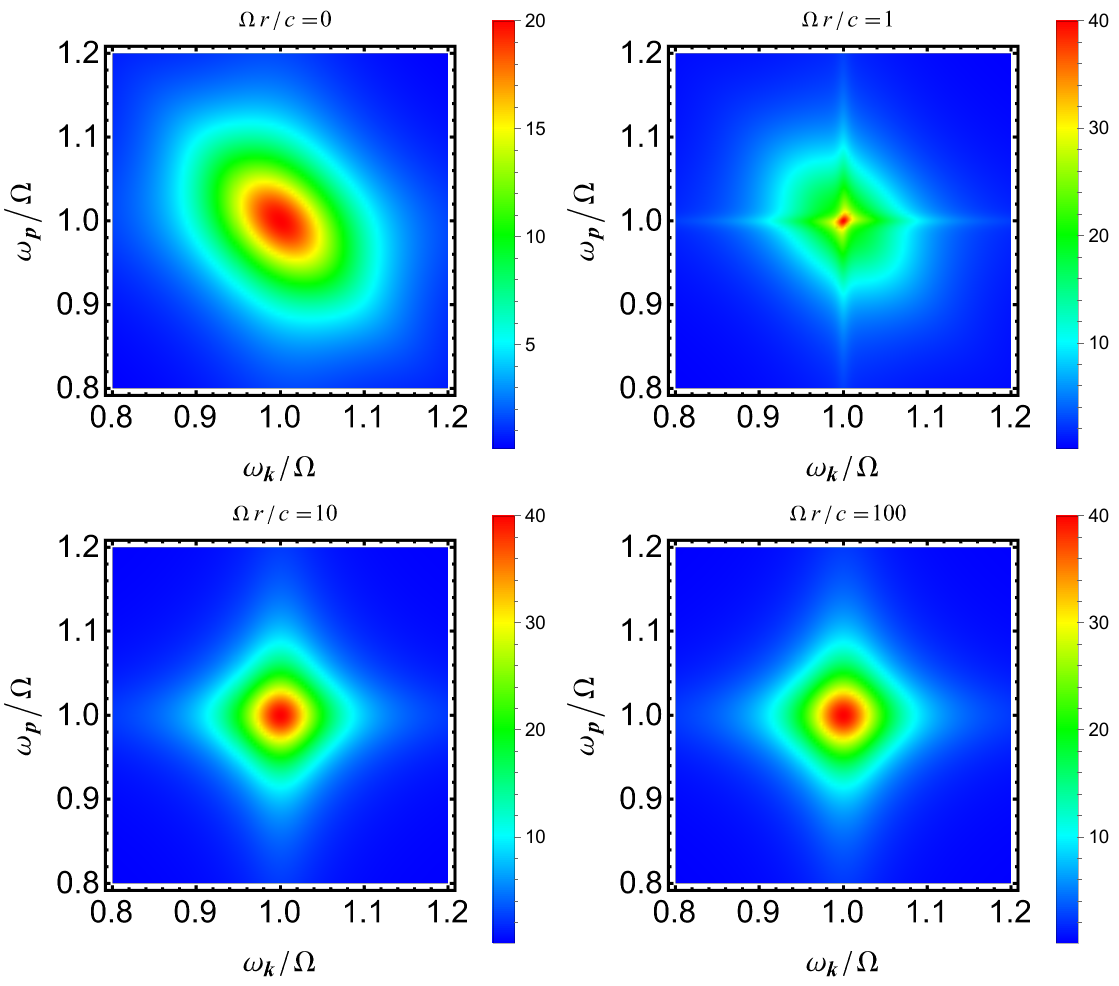}    
    \captionsetup{
  format=plain,
  margin=1em,
  justification=raggedright,
  }
    \caption{Density plots of the two-photon spectral density $\rho(\omega_{\mathbf{k}},\omega_{\mathbf{p}})$ as a function of the photon frequencies $\omega_{\vb k}$ and $\omega_{\vb p}$ for two-atom system. We set $\Gamma/\Omega=0.1$ and vary the atomic separation $r$ from $0$ to $100 c/\Omega$. }
    \label{fig:Two_Photon_Joint_Probability_Density_Plot}
\end{figure}

Density plots of the spectral density $\rho(\omega_{\mathbf{k}},\omega_{\mathbf{p}})$ are shown in Fig. \ref{fig:Two_Photon_Joint_Probability_Density_Plot}. The plots exhibit symmetry with respect to the interchange of $\omega_{\mathbf{k}}$ and $\omega_{\mathbf{p}}$, reflecting the bosonic nature of the two-photon state. Additionally, $\rho(\omega_{\mathbf{k}},\omega_{\mathbf{p}})$ peaks along the line $\omega_{\mathbf{k}}+\omega_{\mathbf{p}}=2\Omega$, implying strong photon correlations in the frequency domain, which arise due to energy conservation. Note that, as may be expected, as the atomic separation increases, the two-photon spectrum evolves into two independent Lorentzian lines, indicating a weakening of photon-photon correlations.

\subsection{\label{sec:IVB}Superradiance and Subradiance}
\begin{figure}[t]
\centering
\includegraphics[width = 0.75\textwidth]{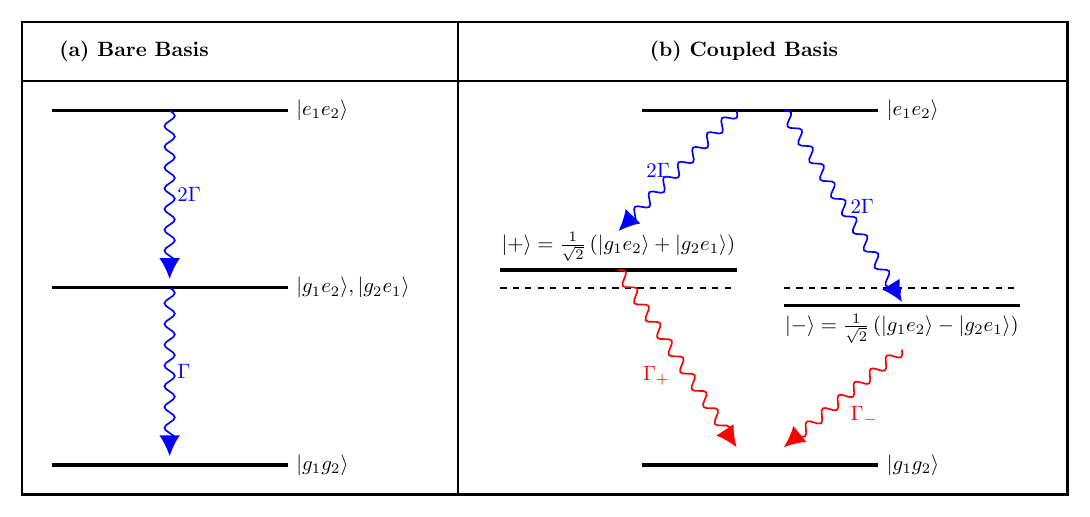}    
   \captionsetup{
  format=plain,
  margin=1em,
  justification=raggedright,
  }
\caption{Illustrating the collective states $\ket{\pm}$ in the two-atom system, assuming $k_{0}r< \pi$.}
\label{twoAtomLevel}
\end{figure}

We now consider the effects of collective emission in the two-atom system. To this end, we introduce the following symmetric and antisymmetric combinations of the mixed-state amplitudes:
\begin{subequations}
    \begin{align}
    b_{+\mathbf{k}}(t) &= \frac{1}{\sqrt{2}} \left(b_{1\mathbf{k}}(t)+b_{2\mathbf{k}}(t)\right) =\frac{1}{\sqrt{2}}  g \frac{e^{-i \mathbf{k} \cdot \mathbf{r}_1} + e^{-i \mathbf{k} \cdot \mathbf{r}_2} }{ \Omega - \omega_{\mathbf{k}} - i\Gamma_{-}/2 } \left(e^{-2i\Omega t - \Gamma t} - e^{-i(\Omega + \omega_{\mathbf{k}})t - \Gamma_{+}t/2} \right), \\
    b_{-\mathbf{k}}(t) &= \frac{1}{\sqrt{2}} \left(b_{1\mathbf{k}}(t)-b_{2\mathbf{k}}(t)\right) = \frac{1}{\sqrt{2}} g \frac{e^{-i \mathbf{k} \cdot \mathbf{r}_2} - e^{-i \mathbf{k} \cdot \mathbf{r}_1} }{ \Omega - \omega_{\mathbf{k}} - i\Gamma_{+}/2 } \left(e^{-2i\Omega t - \Gamma t} - e^{-i(\Omega + \omega_{\mathbf{k}})t - \Gamma_{-}t/2} \right).  
\end{align}
\end{subequations}
The physical meaning of the states $\ket{\pm}$ with amplitudes $b_{\pm\mathbf{k}}(t)$ is illustrated in Fig.~\ref{twoAtomLevel}. The states describe the collective excitations of the atoms due to coupling to the electromagnetic field. The excited state $\ket{e_1 e_2}$ decays to the intermediate states $\ket{\pm}$ at the rate $2\Gamma$. The states $\ket{\pm}$ further decay to the ground state $\ket{g_1 g_2}$ at the rates $\Gamma_{\pm}$, respectively. We note that the relative size of $\Gamma_\pm$ depends upon the separation between the atoms. If $k_0 r < \pi$, then $\Gamma_{+} > \Gamma_{-}$ and the state $\ket{+}$ decays faster than $\ket{-}$.  In this case, $\ket{+}$ is referred to as a superradiant state and $\ket{-}$ as a subradiant state.

Next, to investigate the mode-independent probabilities of the superradiant and subradiant states, we define:
\begin{subequations}
\begin{align}
    &|b_+(t)|^2 = \sum\limits_{\mathbf{k}} \left|b_{+\mathbf{k}}(t) \right|^2  
    = \frac{\Gamma_{+}}{\Gamma_{-}} (e^{-\Gamma_{+} t} - e^{-2 \Gamma t}), \\
   &|b_-(t)|^2=\sum\limits_{\mathbf{k}} \left|b_{-\mathbf{k}}(t) \right|^2  
    = \frac{\Gamma_{-}}{\Gamma_{+}} (e^{-\Gamma_{-} t} - e^{-2 \Gamma t}).
    \label{eq:expression of superradiance and subradiance probability} 
\end{align}
\end{subequations}
Here we have replaced the summation over modes by integrals in the usual manner.

\begin{figure}[t]
    \centering
     \captionsetup{
  format=plain,
  margin=1em,
  justification=raggedright,
  }
    \includegraphics[width =0.75\textwidth]{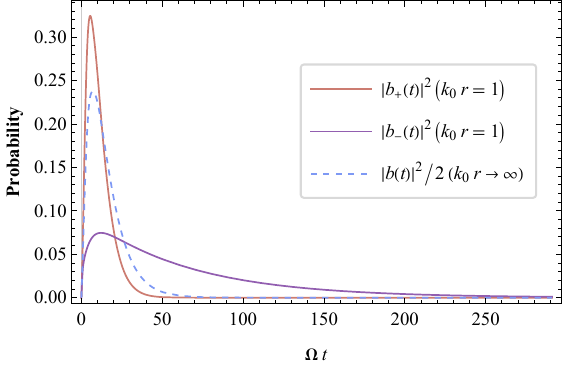}
    \caption {Time dependence of the probabilities of superradiant and subradiant states when the atoms are separated by $r=1/k_0$ and $r\to\infty$. Here we have set $\Gamma/\Omega=0.1$.}
    \label{Two_Photon_sum_of_Superradiance_Subradiance_Plot}
\end{figure}

The time dependence of the quantities $|b_\pm(t)|^2$ are shown in Fig. \ref{Two_Photon_sum_of_Superradiance_Subradiance_Plot}.
For comparison, we also plot $|b_{\pm}(t)|^2$ in the limit when the atomic separation $r \to \infty$. In this case, both the superradiance and subradiance probabilities take one-half of the value of the total probability $|b(t)|^2$, as is evident from Eq.~\eqref{eq:expression of superradiance and subradiance probability}. We observe that initially, both the superradiant and subradiant states are unoccupied. However, when the initial state decays, both the superradiant and subradiant states become populated. The superradiant state reaches a higher maximum value than the subradiant state. At long times, the subradiant state decays more slowly than the superradiant state. This indicates the formation of a so-called dark state at intermediate times~\cite{gegg2018superradiant}.


\subsection{Radiated Power}
To further analyze the collective emission, we calculated the radiated power $P_\pm(t) = \hbar \Omega d \left\vert b_{\pm}(t) \right\vert^{2} / dt$ carried by the superradiant and subradiant states, and the power $P_{c}(t) = 2\hbar \Omega d\left\vert c(t) \right\vert^{2} / dt$ carried by the two-photon state:
\begin{subequations}
    \begin{align}
         P_{+}(t) & = \hbar \Omega \frac{\Gamma_{+}}{\Gamma_{-}} (2\Gamma e^{-2 \Gamma t} - \Gamma_{+} e^{-\Gamma_{+}t}),\\
          P_{-}(t) & = \hbar \Omega \frac{\Gamma_{-}}{\Gamma_{+}} (2\Gamma e^{-2 \Gamma t} - \Gamma_{-} e^{-\Gamma_{-}t}),\\
        P_{c}(t) & = 2\hbar \Omega \left[ 2 \Gamma \left( 1 - \frac{ \Gamma_{+}^{2} + \Gamma_{-}^{2}}{ \Gamma_{+} \Gamma_{-}} \right) e^{-2 \Gamma t} + \frac{\Gamma_{+}^{2}}{\Gamma_{-}} e^{- \Gamma_{+}t}  + \frac{\Gamma_{-}^{2}}{\Gamma_{+}} e^{- \Gamma_{-}t}  \right].
    \end{align}
\label{eq:Ppm}
\end{subequations}
It follows that the total radiated power  $P(t) = P_{+}(t) + P_{-}(t) + P_{c}(t)$ is given by
\begin{equation}
    P(t) = -2\hbar \Omega \Gamma \frac{(\Gamma_{+}- \Gamma_{-})^{2}}{\Gamma_{+} \Gamma_{-}} e^{- 2 \Gamma t } + \hbar \Omega \frac{\Gamma_{+}^{2}}{ \Gamma_{-}} e^{- \Gamma_{+} t} + \hbar \Omega \frac{\Gamma_{-}^{2}}{ \Gamma_{+}} e^{- \Gamma_{-} t}.
    \label{eq:two atom case radiance power}
\end{equation}

\begin{figure}
    \centering
    \captionsetup{
  format=plain,
  margin=1em,
  justification=raggedright,
  }
    \includegraphics[width = 0.7\textwidth]{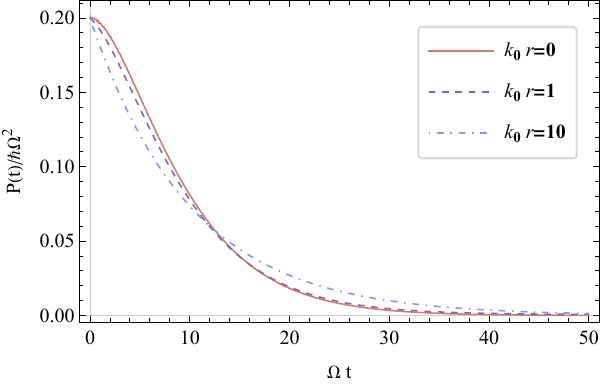}
    \caption{Time dependence of the radiated power for two atom case. The separation between two atoms vary from $r=0$ to $r=10/k_{0}$. Other parameters are chosen to be the same as  in Fig.~\ref{fig:Two_Photon_all_probability_plot}.}
    \label{fig:two atom case radiance power}
\end{figure}

The time-dependence of the radiated power is shown in Fig.\ref{fig:two atom case radiance power}. The power reaches its maximum value $2\hbar \Omega \Gamma$ at $t=0$  and decays monotonically with time afterwards. For the limiting cases $k_{0} r \ll 1$ and $k_{0}r \gg 1$, Eq. \eqref{eq:two atom case radiance power} becomes
\begin{subequations}
    \begin{align}
        &\lim_{k_{0}r\to 0} P(t) = 2 \hbar \Omega \Gamma e^{-2 \Gamma t } (1+ 2 \Gamma t ),\label{eq:total radiance power r is 0}\\
        &\lim_{k_{0}r\to \infty} P(t) = 2 \hbar \Omega \Gamma e^{- \Gamma t }. \label{eq:total radiance power r is large}
    \end{align}
\end{subequations}
We note that Eq. \eqref{eq:total radiance power r is 0} agrees with the phenomenological theory reported in Ref. \cite{gross1982superradiance}. In that work, the authors assumed that the states of the system are symmetric under exchange of atomic positions and employed the principles of probability conservation and energy conservation to derive Eq.~\eqref{eq:total radiance power r is 0}. This assumption holds strictly when the two atoms occupy the same spatial point, i.e. $k_{0}r \to  0$. Specifically, when $k_{0}r = 0$, only the amplitude of the superradiant state $b_{+ \mathbf{k}}$ is non-zero, which is the symmetric combination of $b_{1\mathbf{k}}$ and $b_{2\mathbf{k}}$.

In contrast to Ref. \cite{gross1982superradiance}, we arrive at the same result by means of a first-principle calculation. Additionally, we obtained a more general form of the radiated power of a two-atom system, whose distance dependence is given in Eq.~\eqref{eq:two atom case radiance power}. Moreover for $r \to \infty$, we note that Eq. \eqref{eq:total radiance power r is large} corresponds to twice the radiated power of a single atom, consistent with the fact that two distant atoms radiate as independent emitters.

Finally, we compare the radiated power for the cases $k_{0} r \ll 1$ (red solid line) and $k_{0} r \gg  1$ (blue dot-dashed line) shown in Fig.~\ref{fig:two atom case radiance power}. Initially, the radiated power of two closely spaced atoms is higher than that of two widely separated atoms, but it also decreases more rapidly. This indicates that, as two atoms are brought closer together, they emit energy more quickly, exemplifying superradiant behavior. Such phenomena have been experimentally observed, as reported in Ref.~\cite{mlynek2014observation}. 

\section{\label{sec:small system}Small Systems}
Until now, we have focused on systems comprised of a relatively small number of atoms. In this section, we turn our attention to a system consisting of a constant density of atoms contained in a spherical volume of radius $R$. We study separately the cases of small and large volumes, where $k_0 R \ll 1$ and $k_0 R \gg 1$, respectively. Here $k_0 = \Omega / c$ is the wavenumber corresponding to the atomic transition frequency. In both cases, the analysis begins with the equations of motion Eqs.~\eqref{eq:equation of motion}.

For $k_0 R \ll 1$, the spatial variation of the field can be neglected, allowing us to set the atomic phase factors $e^{i \mathbf{k} \cdot \mathbf{r}_j} = 1$ in Eqs.~\eqref{eq:equation of motion}. With this simplification and applying the constraints in Eqs.~\eqref{eq: constraint of the state}, the equations of motion reduce to:
\begin{subequations}
    \begin{align}
       i \frac{d}{dt} a_{ij}(t) &= 2\Omega a_{ij}(t) + \frac{g}{2} \sum\limits_{\mathbf{k}} \left(b_{i \mathbf{k}}(t) + b_{j \mathbf{k}}(t)\right) - \delta_{ij} g \sum\limits_{\mathbf{k}} b_{i \mathbf{k}}(t), \\
       i \frac{d}{dt} b_{i \mathbf{k}}(t) &= \left( \Omega+\omega_{\mathbf{k}} \right)b_{i \mathbf{k}}(t) + 2g \sum\limits_{j} a_{ij}(t) + 2g \sum\limits_{\mathbf{p}} c_{\mathbf{k}\mathbf{p}}(t), \\
       i \frac{d}{dt} c_{\mathbf{k}\mathbf{p}}(t) &= (\omega_{\mathbf{k}}+\omega_{\mathbf{p}})c_{\mathbf{k}\mathbf{p}}(t) + \frac{g}{2} \sum\limits_{i} \left(b_{i \mathbf{k}}(t) + b_{i \mathbf{p}}(t)\right).
    \end{align}
\end{subequations}

We impose the initial conditions $b_{i\mathbf{k}}(0) = 0$, $c_{\mathbf{kp}}(0) = 0$, and $a_{ij}(0) = 1 / \sqrt{2N(N-1)}$ for $i \neq j$, ensuring that all atoms have equal probability of being excited, consistent with the small size of the system. Taking the Laplace transform of the equations, we obtain:
\begin{subequations}
    \begin{align}
       &i\left( s a_{ij}(s) - a_{ij}(0)\right) = 2\Omega a_{ij}(s) + \frac{g}{2} \sum\limits_{\mathbf{k}} \left(b_{i \mathbf{k}}(s) + b_{j \mathbf{k}}(s)\right) - \delta_{ij} g \sum\limits_{\mathbf{k}} b_{i \mathbf{k}}(s), \label{eq:asmall} \\
       &is b_{i \mathbf{k}}(s) = \left( \Omega+\omega_{\mathbf{k}} \right)b_{i \mathbf{k}}(s) + 2g \sum\limits_{j} a_{ij}(s) + 2g \sum\limits_{\mathbf{p}} c_{\mathbf{k}\mathbf{p}}(s),\label{eq:bsmall} \\
       &is c_{\mathbf{k}\mathbf{p}}(s) = (\omega_{\mathbf{k}}+\omega_{\mathbf{p}})c_{\mathbf{k}\mathbf{p}}(s) + \frac{g}{2} \sum\limits_{i} \left(b_{i \mathbf{k}}(s) + b_{i \mathbf{p}}(s)\right).
       \label{eq:csmall}
    \end{align}
\end{subequations}
 After eliminating $b_{i\mathbf k}$ in Eq. \eqref{eq:asmall}, we find that in the weak-coupling regime, $a_{ij}$ obeys
\begin{align}
(is - 2\Omega)a_{ij}(s) = i a_{ij}(0) +  (1-\delta_{ij}) \Sigma(s,\Omega)  \sum_{l}\left(a_{il}(s) + a_{jl}(s) \right) + \mathcal{O}(g^{4}) , 
\label{eq:aij}
\end{align}
where the definition of the self-energy $\Sigma(s,\Omega)$ is same as Eq. \eqref{expression of self-energy}. Next, we make the pole approximation in the usual manner by replacing $\Sigma(s,\Omega)$ with $\Sigma = \delta \omega - i \Gamma / 2$. Solving Eq.~\eqref{eq:aij} for $a_{ij}$, we obtain
\begin{equation}
    a_{ij}(s) = \frac{i}{\sqrt{2N(N-1)}} \frac{1}{is -2\Omega-2(N-1)\Sigma}+\mathcal{O}(g^{4}).
    \label{eq:solution to N atom for atomic amplitude}
\end{equation}
We note that $a_{ij}$ does not explicitly depend on the indices $i$ and $j$, in accordance with the fact that all atoms are excited with equal probability initially. The detailed derivation of this result is presented in Appendix~\ref{appd:small system}.

We can now obtain the expressions for the amplitudes $b_{i\mathbf{k}}$ and $c_{\mathbf{kp}}$. It follows from Eq.~\eqref{eq:bsmall}, by eliminating the amplitude $c_{\mathbf{kp}}$, that to leading order in $g$, $b_{i \mathbf{k}}$ obeys the equation
\begin{equation}
    \left( is-\Omega-\omega_{\mathbf{k}}-\Sigma(s,\omega_{\mathbf{k}})  \right)b_{i \mathbf{k}}(s)-\Sigma(s,\omega_{\mathbf{k}}) \sum\limits_{j \neq i} b_{j \mathbf{k}}(s) = 2 g \sum\limits_{j}a_{ij}(s) +\mathcal{O}(g^{3}).
\end{equation}
Inserting the formula for $a_{ij}$ from Eq.~\eqref{eq:solution to N atom for atomic amplitude} into the above, making the pole approximation, and solving for $ b_{i \mathbf{k}}$ we find that
\begin{equation}
    b_{i \mathbf{k}}(s) = i g \sqrt{\frac{2(N-1)}{N}} \frac{1}{\left( is-\Omega-\omega_{\mathbf{k}}-N \Sigma \right)(is -2\Omega-2(N-1)\Sigma)}+\mathcal{O}(g^{3}) .
    \label{eq:solution to b in laplace for N atom}
\end{equation}
Continuing in the same manner, we solve Eq.~\eqref{eq:csmall} for $c_{\mathbf{kp}}$ to obtain
\begin{equation}
    c_{\mathbf{k}\mathbf{p}}(s) = \frac{ig^{2}\sqrt{N(N-1) / 2}}{\left( is-\omega_{\mathbf{k}}-\omega_{\mathbf{p}} \right)\left( is-2\Omega-2(N-1)\Sigma \right)\left( is-\Omega-\omega_{\mathbf{k}}-N\Sigma \right)}+(\mathbf{k} \leftrightarrow \mathbf{p})+\mathcal{O}(g^{4}).
\end{equation}        

Finally, inverting the Laplace transforms, the mode- and atom-independent probabilities $|a(t)|^2$, $|b(t)|^2$ and $|c(t)|^2$  are given by 
\begin{subequations}
    \begin{align}
        &\left\vert a(t) \right\vert^{2}  = e^{-2(N-1)\Gamma t },\\
        & \left\vert b(t) \right\vert^{2} = 2\frac{N-1}{N-2} \left[ e^{-N \Gamma t } - e^{- 2(N-1) \Gamma t } \right],\\
        & \left\vert c(t) \right\vert^{2} = 1- 2 \frac{N-1}{N-2} e^{-N \Gamma t } + \frac{N}{N-2} e^{-2(N-1) \Gamma t } .
    \end{align}
\label{eq:N_atom_probabilities}
\end{subequations}
As usual, we have replaced the sums over modes by integrals and regularized the divergences in the integrals. The derivation details are presented in Appendix~\ref{appd:small system}.

\begin{figure}[htbp]
    \centering
    \captionsetup{
  format=plain,
  margin=1em,
  justification=raggedright,
  }
    \includegraphics[width = 0.75\textwidth]{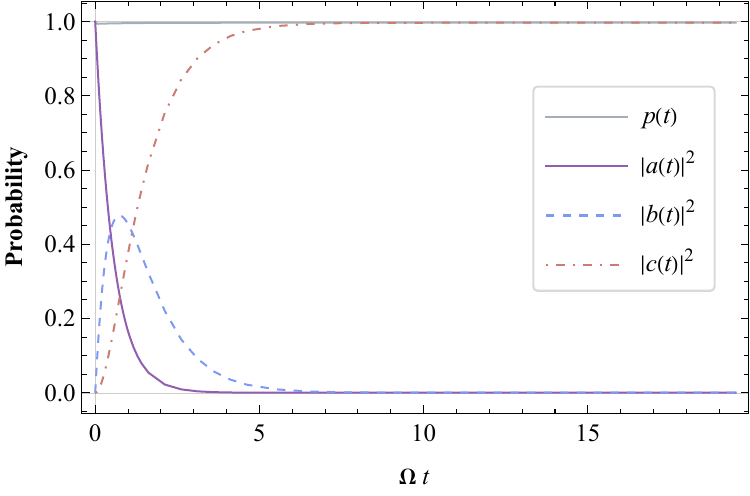}
    \caption{Time dependence of the probabilities $|a(t)|^2$, $|b(t)|^2$, $|c(t)|^2$ and $p(t)$ for a small system with $N=10$ atoms and $\Gamma / \Omega= 0.1$.}
    \label{fig:N atom case all probability plot}
\end{figure}

The time evolution of the probabilities $|a(t)|^2$, $|b(t)|^2$, and $|c(t)|^2$ is shown in Fig.~\ref{fig:N atom case all probability plot} for a system with $N=10$ atoms. Additionally, the total probability $p(t) = |a(t)|^2 + |b(t)|^2 + |c(t)|^2$ is plotted, demonstrating conservation over time. The results exhibit a similar qualitative behavior to the two-atom case shown in Fig.~\ref{fig:Two_Photon_all_probability_plot}, with the two-atom excitation probability $|a(t)|^2$ decaying exponentially, the mixed-state probability $|b(t)|^2$ peaking before decreasing, and the two-photon probability $|c(t)|^2$ asymptotically approaching unity. Notably, the decay rate is approximately $N$ times larger than in the single-atom case, as the self-energy correction scales with $N-1$, as evident from Eq.~\eqref{eq:N_atom_probabilities}.

\subsection{Two-Photon Spectrum}
We now consider the two-photon spectrum, following the ideas of Section~\ref{spectrum two atom}. We begin by noting that at long times, the total probability $p(t)$ is of the form
\begin{equation}
   p(t\to\infty) = \left\vert c(t\to \infty) \right\vert^{2}  = \int_{-\infty}^{\infty} \int_{-\infty}^{\infty} d \omega_{\mathbf{k}} d \omega_{\mathbf{p}} \rho(\omega_{\mathbf{k}},\omega_{\mathbf{p}}),
    \label{eq:spectral_integral n atom}
\end{equation}
where the two-photon spectral density is given by
\begin{equation}
    \rho(\omega_{\mathbf{k}},\omega_{\mathbf{p}}) = \frac{ N(N-1) \Gamma^{2} \left(\left( 2\Omega-\omega_{\mathbf{k}}-\omega_{\mathbf{p}} \right)^{2}+N^{2}\Gamma^{2} \right) / 4 \pi^{2} }{ \left( (2\Omega-\omega_{\mathbf{k}}-\omega_{\mathbf{p}})^{2}+(N-1)^{2}\Gamma^{2} \right)\left( \left( \Omega-\omega_{\mathbf{k}} \right)^{2}+N^{2}\Gamma^{2} / 4 \right)\left( \left( \Omega-\omega_{\mathbf{p}} \right)^{2}+N^{2}\Gamma^{2} / 4 \right)}.
    \label{eq:photon spectrum for the N atom case}
\end{equation}

\begin{figure}[t]
     \centering
     \captionsetup{
  format=plain,
  margin=1em,
  justification=raggedright,
  }
    \includegraphics[width = 0.75\textwidth]{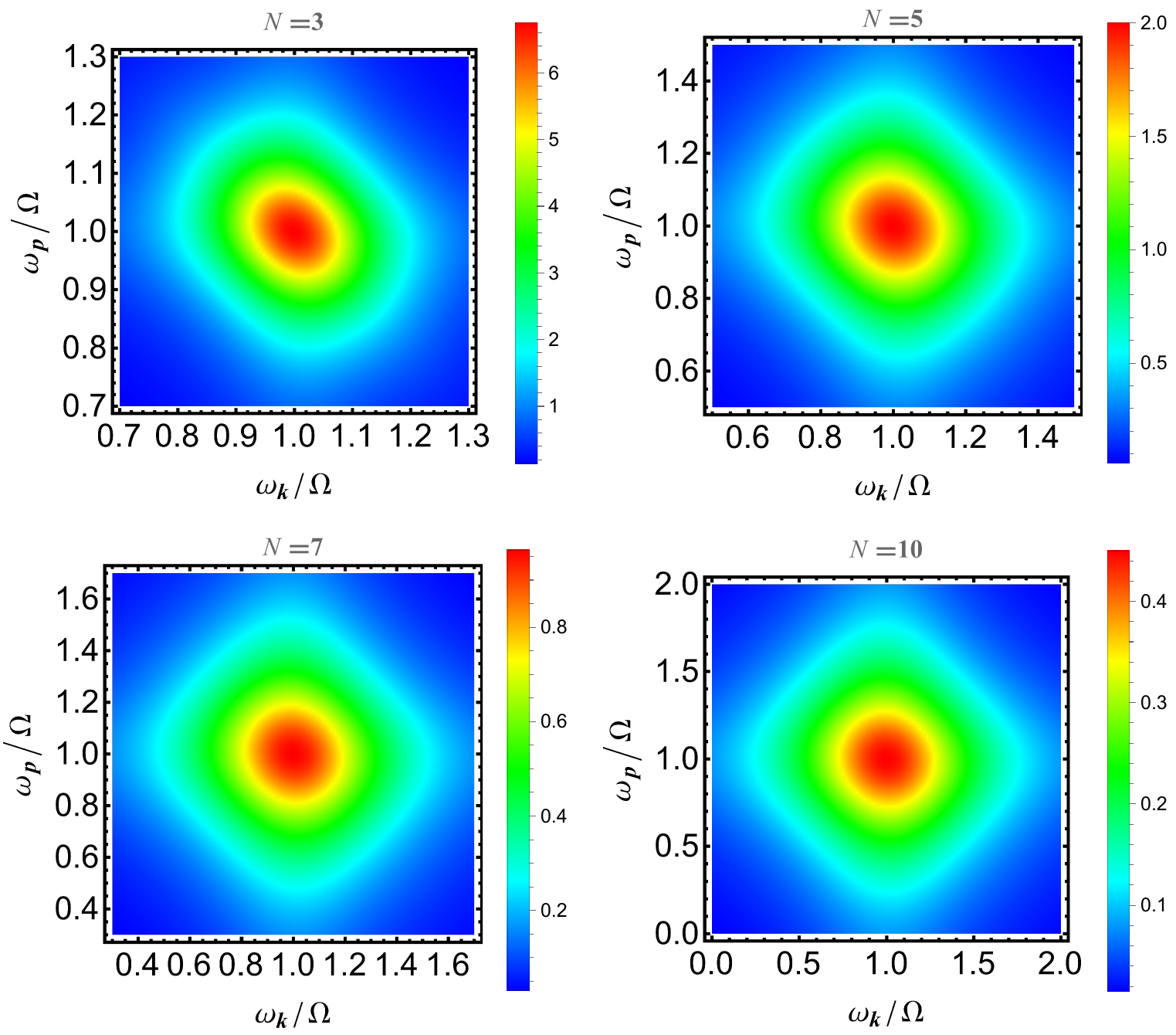}
    \caption{Two-photon spectrum for small atomic systems with $N= 3, 5, 7, 10$ atoms and
    $\Gamma = 0.1\Omega$. }
    \label{fig:N atom case photon spectrum}
\end{figure}

Density plots of $\rho(\omega_{\mathbf{k}},\omega_{\mathbf{p}})$ are shown in Fig.~\ref{fig:N atom case photon spectrum} for various values of the number of atoms $N$. The photon spectrum exhibits a pronounced peak along the line $\omega_{\mathbf{k}}+\omega_{\mathbf{p}}=2\Omega$, reflecting strong photon-photon correlations due to energy conservation. As $N$ increases, $\rho(\omega_{\mathbf{k}},\omega_{\mathbf{p}})$ gradually transitions into two independent Lorentzian spectral lines, each determined  by the frequencies $\omega_{\mathbf{k}}$ and $\omega_{\mathbf{p}}$. This transition indicates that photon-photon correlations weaken in the systems with large number of atoms, resulting in uncorrelated photon emissions.

\subsection{Radiated Power}
We now consider the radiated power, following the approach in Sec. \ref{sec:IVB}. In the case of the mixed state, with $ P_{b}(t) = \hbar \Omega ~ d \left\vert b(t) \right\vert^{2} /  dt$ and the two-photon state, with $ P_{c}(t) = 2 \hbar \Omega ~ d \left\vert c(t) \right\vert^{2} /  dt$, we obtain
\begin{subequations}
    \begin{align}
        P_{b}(t) &=  2 \hbar \Omega \Gamma \frac{N-1}{N-2} \left[ 2(N-1) e^{-2(N-1)\Gamma t} - N e^{-N \Gamma t } \right],\\
        P_{c}(t)  &=  4 \hbar \Omega \Gamma \frac{N(N-1)}{N-2} (e^{- N  \Gamma t } - e^{-2(N-1) \Gamma t }).
    \end{align}
\end{subequations}
The total radiated power $P(t) = P_{b}(t) + P_{c}(t)$ is given by
\begin{equation}
    P(t) = 2 \hbar \Omega \Gamma \frac{N-1}{N-2} \left[ N e^{- N \Gamma t } - 2 e^{-2 (N-1) \Gamma t }\right].
    \label{eq:radiance power of N atom}
\end{equation}

\begin{figure}[t]
    \centering
    \captionsetup{
  format=plain,
  margin=1em,
  justification=raggedright,
  }
    \includegraphics[width = 0.7\textwidth]{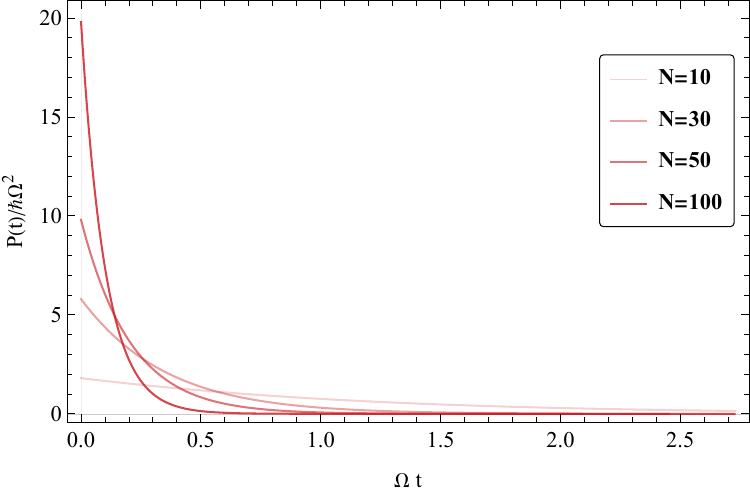}
    \caption{Time dependence of the radiated power for small systems. The parameters are chosen to be the same as  in Fig.~\ref{fig:N atom case photon spectrum}.}
    \label{fig:N atom case small size time dependence and scaling with atom number}
\end{figure}
The time-dependence of the radiated power for systems varying from 10 to 100 atoms in size is shown in Fig.~\ref{fig:N atom case small size time dependence and scaling with atom number}. It can be seen that the power decreases monotonically with time. The maximum power $P_\text{max}$ is achieved at $t=0$ and is given by 
\begin{equation}
    P_{\text{max}} = 2  \hbar \Omega \Gamma (N-1).
    \label{eq:maximum power for N atom case}
\end{equation} 

In contrast to the case of Dicke superradiance, where $P_\text{max} \propto N^2$ ~\cite{brito2020superradiance}, we find that $P_\text{max} \propto N$. This can be explained by the fact that in  the Dicke model, all of the atoms are prepared in their excited states, so the total photon energy is $N \hbar \Omega$. In addition, the atoms decay cooperatively, so that the decay rate is a factor of $N$ larger than the single-atom decay rate $\Gamma$, leading to the $N^2$ scaling of $P_\text{max}$. Here the number of excitations is fixed to be two, independent of $N$. In this setting, the self-energy is proportional to $N$, leading to $P_\text{max}$ proportional to $N$, rather than $N^2$. In particular, when $N \gg 1$, Eq.~\eqref{eq:radiance power of N atom} becomes
\begin{equation}
P(t ) = 2 \hbar \Omega  \Gamma N e^{-N\Gamma t}
\label{eq:radiated power of N is large}
\end{equation}
Eq.~\eqref{eq:radiated power of N is large} reveals that when all atoms are in phase, they collectively behave as a ``giant atom", spontaneously emitting radiation like a single atom. However, the decay rate is enhanced by a factor of $N$, resulting in a purely exponential decay rate that is $N$ times larger than the spontaneous emission rate of an individual atom.

\section{Large Systems}
\label{sec:large systems}
We now consider the case of large atomic systems, where $k_0 R \gg 1$. In this limit, we can no longer ignore the spatial variation of the field. We assume that the atoms are uniformly distributed at constant density $\rho_0$ in a volume $V_a$. We also assume that the atoms are initially in their ground states. In this setting, we treat the system as continuous and replace all discrete quantities by their continuous counterparts according to $a_{ij}(t) \to a(\mathbf{x},\mathbf{y},t),~ b_{i\mathbf{k}}(t) \to  b(\mathbf{x},\mathbf{k},t)~ \rm{and}~ c_{\mathbf{k}\mathbf{p}}(t) \to  c(\mathbf{k},\mathbf{p},t)$. We also replace the sum over atoms by an integral: $\sum\limits_{i} \to \rho_0\int_{V_a}  d^{3}x $.  The summation over modes is also replaced by an integral in the usual manner. With these modifications  Eqs.~\eqref{eq:equation of motion} becomes for $\mathbf{x} \neq \mathbf{y}$,
\begin{subequations}
    \begin{align}
    i \frac{\partial}{\partial t}a(\mathbf{x},\mathbf{y},t) =& 2\Omega a(\mathbf{x},\mathbf{y},t) + \frac{1}{2}gV   \int \frac{d^{3} k}{(2\pi)^{3}}\left( b(\mathbf{x},\mathbf{k},t)e^{i \mathbf{k}\cdot\mathbf{y}} + b(\mathbf{y},\mathbf{k},t)e^{i \mathbf{k}\cdot\mathbf{x}} \right),\\
    i \frac{\partial}{\partial t}b(\mathbf{x},\mathbf{k},t)=&(\Omega+\omega_{\mathbf{k}})b(\mathbf{x},\mathbf{k},t) \nonumber \\
    & + 2 g  \rho_0 \int_{V_a} d^{3}y \ a(\mathbf{x},\mathbf{y},t) e^{-i\mathbf{k}\cdot\mathbf{y}}+ 2g V \int \frac{d^{3} p}{(2\pi)^3} \ c(\mathbf{k},\mathbf{p},t)e^{i\mathbf{p}\cdot\mathbf{x}},\\
    i \frac{\partial}{\partial t}c(\mathbf{k},\mathbf{p},t) =& (\omega_{\mathbf{k}}+\omega_{\mathbf{p}})c(\mathbf{k},\mathbf{p},t)+ \frac{g}{2} \rho_0 \int_{V_a} d^{3} x  \left(b(\mathbf{x},\mathbf{k},t) e^{-i \mathbf{p} \cdot \mathbf{x}}+b(\mathbf{x},\mathbf{p},t)e^{-i\mathbf{k}\cdot\mathbf{x}} \right) ,
    \end{align}
\label{eq:continuum_limit}
\end{subequations}
and  $a(\mathbf{x, x}) =0$.
Upon Laplace transforming  Eqs.~\eqref{eq:continuum_limit} we obtain
\begin{subequations}
    \begin{align}
        &i( s a(\mathbf{x},\mathbf{y},s)  - a(\mathbf{x},\mathbf{y},0) ) \nonumber \\
        &= 2\Omega a(\mathbf{x},\mathbf{y},s) + \frac{1}{2}gV \int \frac{d^{3} k}{(2\pi)^{3}}\left( b(\mathbf{x},\mathbf{k},s)e^{i \mathbf{k}\cdot\mathbf{y}} + b(\mathbf{y},\mathbf{k},s)e^{i \mathbf{k}\cdot\mathbf{x}} \right),\label{eq:large system Laplace transform for a}\\
        &i s b(\mathbf{x},\mathbf{k},s)=(\Omega+\omega_{\mathbf{k}})b(\mathbf{x},\mathbf{k},s)+ 2 g \rho_0 \int_{V_a} d^{3}y \ a(\mathbf{x},\mathbf{y},s) e^{-i\mathbf{k}\cdot\mathbf{y}}+  2 g V \int \frac{d^{3} p}{(2\pi)^3} \ c(\mathbf{k},\mathbf{p},t)e^{i\mathbf{p}\cdot\mathbf{x}},\label{eq:large system Laplace transform for b}\\
        &i s c(\mathbf{k},\mathbf{p},s) = (\omega_{\mathbf{k}}+\omega_{\mathbf{p}})c(\mathbf{k},\mathbf{p},s)+ \frac{g}{2} \rho_0 \int_{V_a} d^{3} x  \left(b(\mathbf{x},\mathbf{k},s) e^{-i \mathbf{p} \cdot \mathbf{x}}+b(\mathbf{x},\mathbf{p},s)e^{-i\mathbf{k}\cdot\mathbf{x}} \right) .
        \label{eq:large system Laplace transform for c}
    \end{align}
\label{eq:large system Laplace transform}
\end{subequations}
Here we have imposed the initial conditions $b_{i\mathbf{k}}(0)=0$ and $c_{\mathbf{kp}}(0)=0$, so that no photons are present in the field initially.

Next, we eliminate $b(\mathbf{x},\mathbf{k},s)$ from Eqs.~\eqref{eq:large system Laplace transform}. We find that in the weak-coupling regime ($g / \Omega \ll 1$) and after making the pole approximation, $a(\mathbf{x},\mathbf{y},s)$ obeys
\begin{equation}
    \begin{aligned}
    &(is - 2\Omega) a(\mathbf{x},\mathbf{y},s) = i a(\mathbf{x},\mathbf{y},0) + \rho_{0} \int_{V_{a}} d^{3} {z} \left[ \Delta(\mathbf{y}-\mathbf{z}) a(\mathbf{z},\mathbf{x},s) +  (\mathbf{x} \leftrightarrow \mathbf{y}) \right] + \mathcal{O}(g^{4}).
    \end{aligned}
    \label{eq:equation for a in lagre system}
\end{equation}
Here $\Delta(\mathbf{x})$ is defined by 
\begin{equation}
    \Delta(\mathbf{x}) = g^2 V\int \frac{d^{3} {k}}{(2\pi)^{3}} \frac{e^{i \mathbf{k}\cdot \mathbf{x}}}{\Omega - \omega_{\mathbf{k}} + i \epsilon} = \delta \omega (\left\vert \mathbf{x}\right\vert )- i \frac{\Gamma}{2} \text{sinc} (k_{0} \left\vert \mathbf{x} \right\vert ) ,
\end{equation}
where $k_0=\Omega/c$ and $\epsilon>0$ is small.
The quantity $\Gamma$ is the single-atom rate of spontaneous emission as defined by Eq.~\eqref{def_Gamma} and $\delta \omega$ is the Lamb shift, which we will subsequently neglect. 

Eq. \ref{eq:equation for a in lagre system} is an integral equation for $a(\mathbf{x},\mathbf{y},s)$. The equation can be solved by expanding the solution in eigenfunctions of a suitable operator.   We begin by observing that $\text{sinc}(k_{0} \left\vert \mathbf{x} - \mathbf{y} \right\vert)$ can be written in the form \cite{Abramowitz_Stegun}
\begin{equation}
    \text{sinc}(k_{0}\left\vert \mathbf{x}-\mathbf{y} \right\vert ) = 4 \pi \sum\limits_{l=0}^{\infty} \sum\limits_{m=-l}^{l} j_{l}(k_{0}x) j_{l}(k_{0}y) Y_{lm}(\mathbf{\hat{x}})Y_{lm}^{*}(\mathbf{\hat{y}}).
\end{equation}
This result suggests that we expand $a(\mathbf{x},\mathbf{y},s)$ as
\begin{equation}
    a(\mathbf{x},\mathbf{y},s) =  \sum\limits_{l=0}^{\infty} \sum\limits_{m=-l}^{l} a_{lm}(s) j_{l}(k_{0}x) j_{l}(k_{0}y) Y_{lm}(\mathbf{\hat{x}})Y_{lm}^{*}(\mathbf{\hat{y}}).
    \label{eq:expansion of a}
\end{equation}
Note that Eq. \eqref{eq:expansion of a} respects the bosonic symmetry of $a(\mathbf{x},\mathbf{y},s)$. To find the coefficients $a_{lm}(s)$, we substitute  Eq.~\eqref{eq:expansion of a} into  Eq. \eqref{eq:equation for a in lagre system} and use the orthogonality of the spherical harmonics to obtain
\begin{equation}
    a_{lm}(s) =  \frac{i a_{lm}(0)}{is - 2\Omega + i \lambda_{l} \Gamma} .
    \label{eq:large system solution for a}
\end{equation}
The eigenvalues $\lambda_{l}$ are given by
\begin{equation}
    \lambda_{l} = 4 \pi \rho_{0} \int_{0}^{R} dr ~ r^{2} j_{l}^{2}(k_{0} r) = \frac{3}{2} N (j_{l}^{2}(k_{0}R) - j_{l-1}(k_{0}R) j_{l+1}(k_{0}R)),
    \label{eq:lambda l}
\end{equation}
where $N = 4 \pi \rho_{0} R^{3} / 3$ is the number of atoms in the volume $V_a$. 
We note that when $k_0 R \gg l$, $j_{l}(x) \sim \sin{(x-l\pi /2)} /x $. Consequently, we obtain the asymptotic form of Eq. \eqref{eq:lambda l} as
\begin{equation}
\lambda_l \sim \frac{3N}{2(k_0R)^2} .
\end{equation}

The details of the calculation of Eq. \eqref{eq:large system solution for a} are given in Appendix \ref{appd:continuum}. Continuing as above, we find that $b(\mathbf{x},\mathbf{k},s)$ obeys
\begin{equation}
    (is -\Omega - \omega_{\mathbf{k}})b(\mathbf{x},\mathbf{k},s) = 2 g \rho_{0} \int_{V_{a}} d^{3} {y} a(\mathbf{x},\mathbf{y},s) e^{- i \mathbf{k} \cdot \mathbf{y}} + \rho_{0} \int_{V_{a}} d^{3} {y} \Delta(\mathbf{x},\mathbf{y}) b(\mathbf{y},\mathbf{k},s).
\end{equation}
Substituting the expression Eq.~\eqref{eq:large system solution for a} into the above and making use of the result \cite{Abramowitz_Stegun}
\begin{equation}
    e^{-i \mathbf{k} \cdot \mathbf{r}} = 4 \pi \sum\limits_{l=0}^{\infty} \sum\limits_{m=-l}^{l} (-i)^{l} j_{l}(k r) Y_{lm}(\mathbf{\hat{r}}) Y_{lm}^{*}(\mathbf{\hat{k}}),
    \label{eq:plane wave expansion}
\end{equation}
we obtain
\begin{equation}
    b(\mathbf{x},\mathbf{k},s) = -\sum\limits_{l,m} \frac{2 g (-i)^{l+1} \beta_{l}(k) a_{lm}(0) }{( is - 2 \Omega + i \lambda_{l} \Gamma)(is - \Omega - \omega_{\mathbf{k}} + \frac{1}{2} i \lambda_{l} \Gamma)} j_{l}(k_{0}x) Y_{lm}(\mathbf{\hat{x}}) Y_{lm}^{*}(\mathbf{\hat{k}}) .
    \label{eq:large system solution for b}
\end{equation}
Here we have once again used the orthogonality of the spherical harmonics and have defined
\begin{equation}
    \beta_{l}(k) = 4\pi \rho_{0} \int_{0}^{R}dr ~ r^{2} j_{l}(k_{0} r ) j_{l}(k r).
\end{equation}

Substituting Eq. \eqref{eq:plane wave expansion} and Eq. \eqref{eq:large system solution for b} into Eq. \eqref{eq:large system Laplace transform for c}, we find
\begin{equation}
    c(\mathbf{k},\mathbf{p},s) = \frac{1}{2}\sum\limits_{l,m} \frac{i g^{2} (-1)^{l} \beta_{l}(k) \beta_{l}(p) a_{lm}(0) Y_{lm}^{*}(\mathbf{\hat{k}})Y_{lm}(\mathbf{\hat{p}})}{(is - 2 \Omega + i \lambda_{l} \Gamma)(is - \Omega - \omega_{\mathbf{k}} + \frac{1}{2} i \lambda_{l} \Gamma)(is - \omega_{\mathbf{k}} -  \omega_{\mathbf{p}})} + (\mathbf{k} \leftrightarrow \mathbf{p}).
    \label{eq:large system solution for c}
\end{equation}
Using the relations $Y_{lm}^{*}(\mathbf{\hat{r}}) = (-1)^{-m} Y_{l-m}(\mathbf{\hat{r}})$, $a_{lm}(0) = a_{l-m}(0)$ (arising from the symmetry constraint $a(\mathbf{x}, \mathbf{y}, s) = a(\mathbf{y}, \mathbf{x}, s)$) and make approximation $\beta_{l}(k) \approx \beta_{l}(k_{0}) \equiv \lambda_{l}  $ by the on-shell approximation, we express the Eq. \eqref{eq:large system solution for c} in a more compact form as
\begin{equation}
    c(\mathbf{k},\mathbf{p},s) = \sum\limits_{l,m} \frac{i g^{2} (-1)^{l} \lambda_{l}^{2} a_{lm}(0) Y_{lm}^{*}(\mathbf{\hat{k}})Y_{lm}(\mathbf{\hat{p}}) (2is - 2 \Omega- \omega_{\mathbf{k}} - \omega_{\mathbf{p}} + i \lambda_{l} \Gamma)}{(is - 2 \Omega + i \lambda_{l} \Gamma)(is - \omega_{\mathbf{k}} -  \omega_{\mathbf{p}})(is - \Omega - \omega_{\mathbf{k}} + \frac{1}{2} i \lambda_{l} \Gamma) (is - \Omega - \omega_{\mathbf{p}} + \frac{1}{2} i \lambda_{l} \Gamma)} .
    \label{eq:cbig}
\end{equation}

Finally, inverting the Laplace transform, we find that the mode-independent probabilities are given by
\begin{subequations}
    \begin{align}
        \left\vert a(t) \right\vert^{2} & = \frac{1}{8\pi^{2}} \sum_{l,m} \lambda_{l}^{2} \left\vert a_{lm}(0) \right\vert^{2}  e^{-2 \lambda_{l} \Gamma t} ,\\
        \left\vert b(t) \right\vert^{2} &=\frac{1}{4\pi^{2}}
        \sum_{l,m} \lambda_{l}^{2} \left\vert a_{lm}(0) \right\vert^{2} (e^{-\lambda_{l} \Gamma t } - e^{-2 \lambda_{l} \Gamma t }), \\
        \left\vert c(t) \right\vert^{2} &= \frac{1}{8\pi^{2}} \sum\limits_{l,m} \lambda_{l}^{2} \left\vert a_{lm}(0) \right\vert^{2}(1- 2 e^{- \lambda_{l} \Gamma t } + e^{- 2 \lambda_{l} \Gamma t }) .
    \end{align}
    \label{eq:probabilities for large system}
\end{subequations}
Here we evaluate the integral over $\mathbf{k}$ on shell
by taking $\int d^{3} {k} = k_{0}^{2} \int dk d \hat{\mathbf{k}}$, as previously. The details of derivation is given in the Appendix \ref{appd:prob large system}. We note that the total probability in the continuum case is also conserved because
\begin{equation}
    p(t) = \left\vert a(t) \right\vert^{2} + \left\vert b(t) \right\vert^{2} + \left\vert c(t) \right\vert^{2} = \frac{1}{8\pi^{2}} \sum_{l,m} \lambda_{l}^{2} \left\vert a_{lm}(0) \right\vert^{2} = \left\vert a(0) \right\vert^{2} =1,
\end{equation}

\begin{figure}[htbp]
    \centering
    \captionsetup{
  format=plain,
  margin=1em,
  justification=raggedright,
  }
    \includegraphics[width = 0.75\textwidth]{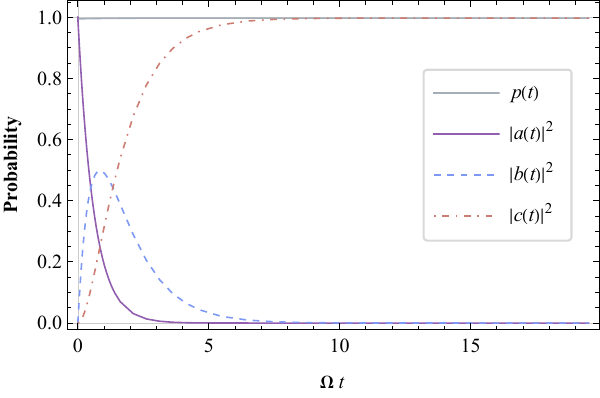}
    \caption{Time dependence of the probabilities $|a(t)|^2$, $|b(t)|^2$, $|c(t)|^2$ and $p(t)$
    for $s-$wave scattering with $N=100,~ k_{0}R=4.0$ and $\Gamma / \Omega= 0.1$.}
    \label{fig:large system all probability plot}
\end{figure}

In Fig. \ref{fig:large system all probability plot}, we illustrate the time evolution of the probabilities for $s$-wave scattering, assuming $a_{00}(0) = 1$ and $a_{lm}(0) = 0$ for $l \geq 1$. The parameters are set to $N = 100$ and $k_{0}R = 4.0$. The dynamics exhibit a familiar pattern: the probability $|a(t)|^2$ decays exponentially, while $|b(t)|^2$ initially rises to a peak before decaying, and $|c(t)|^2$ gradually increases, approaching unity at long times. The total probability $p(t)$ remains conserved throughout the evolution, as expected.


\subsection{Two-Photon Spectrum}
We now compute the two-photon spectrum for large systems. We find the total probability at long time is given by
\begin{equation}
    p(t\to \infty) = \left\vert c(t \to  \infty) \right\vert^{2}  =\int_{-\infty}^{\infty}d\omega_{\mathbf{k}}\int_{-\infty}^{\infty}d\omega_{\mathbf{p}} \rho(\omega_{\mathbf{k}},\omega_{\mathbf{p}}),
\end{equation}
where
\begin{equation}
   \rho(\omega_{\mathbf{k}},\omega_{\mathbf{p}}) = \frac{\Gamma^{2}}{32 \pi^{4}} \sum\limits_{l,m}  \frac{ \lambda_{l}^{4} \left\vert a_{lm}(0) \right\vert^{2} }{\left[ (\Omega - \omega_{\mathbf{k}})^{2} + \frac{1}{4} \lambda_{l}^{2} \Gamma^{2} \right] \left[ (\Omega - \omega_{\mathbf{p}})^{2} + \frac{1}{4} \lambda_{l}^{2} \Gamma^{2} \right]}.
\end{equation}
In Fig. \ref{fig:large system two-photon spectrum}, we plot the spectral density $\rho(\omega_{\mathbf{k}},\omega_{\mathbf{p}})$ using the same parameters as in Fig. \ref{fig:large system all probability plot}. As may be expected, $\rho(\omega_{\mathbf{k}}, \omega_{\mathbf{p}})$ exhibits two independent Lorentzian lines for $s-$wave scattering. 

\subsection{Two-photon Entanglement}
The two-photon spectrum is a measure of photon-photon correlations. A more precise characterization of such correlations is provided by the von Neumann entropy, viewed as a measure of two-photon entanglement. We begin by considering the long-time limit of the state $\ket{\Psi(t)}$, defined in Eq.~\eqref{eq: def of state}, in which only the contribution from the two-photon amplitude $c(\mathbf{k},\mathbf{p},t)$ survives. The corresponding density matrix $\hat{\rho}(t)$ is defined by
\begin{equation}
    \hat{\rho}(t) = \lvert \Psi(t) \rangle \langle \Psi(t) \rvert,
    \label{eq:def density}
\end{equation}
where $ \lvert \Psi (t) \rangle 
= \sqrt{2} \sum_{\mathbf{k}, \mathbf{p}}c(\mathbf{k},\mathbf{p},t) \lvert \mathbf{k} \rangle_{A}  \otimes\lvert \mathbf{p} \rangle_{B} $. Here the photons are distinguished by the labels $A$ and $B$, and the factor of $\sqrt{2}$ is ensures that the state is properly normalized, consistent with Eq.~\eqref{eq:def of mode ind prob}. It follows from  Eq.~\eqref{eq:cbig} that in the long-time limit, the probability amplitude $c(\mathbf{k},\mathbf{p},t)$ is given by
\begin{equation}
    c(\mathbf{k},\mathbf{p},t)  = \sum\limits_{lm} \frac{g^{2} (-1)^{l} \lambda_{l}^{2} a_{lm}(0) e^{- i(\omega_{\mathbf{k}}+ \omega_{\mathbf{p}})t} }{ (\omega_{\mathbf{k}} - \Omega + \frac{1}{2} i \lambda_{l} \Gamma)(\omega_{\mathbf{p}} - \Omega + \frac{1}{2} i \lambda_{l} \Gamma)} Y_{lm}^*(\hat{\mathbf{k}}) Y_{lm}(\hat{\mathbf{p}}).
    \label{eq:clong}
\end{equation}
It will prove to be useful to introduce the bases for the single-photon Hilbert space $\lvert u_{lm}(t) \rangle_{A} = \sum_{\mathbf{k}} \psi^{*}_{lm}(\mathbf{k},t) \lvert \mathbf{k} \rangle_{A} $ and $\lvert v_{lm}(t) \rangle_{B} = \sum_{\mathbf{p}} \phi_{lm}(\mathbf{p},t) \lvert \mathbf{p} \rangle_{B} $. Here the amplitudes $\psi_{lm},~\phi_{lm}$ are defined by
\begin{subequations}
        \begin{align}
            \psi_{lm}(\mathbf{k},t) & = \frac{\sqrt{4 \pi \lambda_{l}} (-i)^{l} g e^{ i \omega_{\mathbf{k}}  t }}{ \omega_{\mathbf{k}} - \Omega - \frac{1}{2} i \lambda_{l} \Gamma} Y_{lm}(\mathbf{\hat{k}}), \\
            \phi_{lm}(\mathbf{p},t) & = \frac{\sqrt{4 \pi \lambda_{l}} i^{l} g e^{ -i \omega_{\mathbf{p}}  t }}{ \omega_{\mathbf{p}} - \Omega + \frac{1}{2} i \lambda_{l} \Gamma} Y_{lm}(\mathbf{\hat{p}}),
        \end{align}
\end{subequations}
Using the orthogonality of spherical harmonics, it is easily verified that the basis states $\lvert u_{lm}(t) \rangle_{A}$ and $\lvert v_{lm}(t) \rangle_{B}$ are orthonormal:
\begin{subequations}
    \begin{align}
        \langle u_{lm}(t) \lvert u_{l^{\prime} m^{\prime} } (t) \rangle_{A} &= \sum_{\mathbf{k}} \psi_{lm}(\mathbf{k},t) \psi_{l'm'}^{*}(\mathbf{k},t) = \delta_{ll'} \delta_{mm'}, \\
        \langle v_{lm}(t) \lvert v_{l^{\prime} m^{\prime} } (t) \rangle_{B} &=  \sum_{\mathbf{p}} \phi_{lm}^{*}(\mathbf{p},t) \phi_{l'm'}(\mathbf{p},t) = \delta_{ll'} \delta_{mm'} ,
    \end{align}
    \label{eq:orthonormal}
\end{subequations}
where the sum is evaluated as an integral within the on-shell approximation.
In the tensor product basis $\lvert u_{lm}(t) \rangle_{A}\otimes\lvert v_{lm}(t) \rangle_{B}$, the two-photon state becomes
\begin{align}
\lvert \Psi (t) \rangle = \sum_{lm} c_{lm} \lvert u_{lm} (t )\rangle_{A} \otimes \lvert v_{lm} (t )\rangle_{B}, 
\end{align}
where $c_{lm} = \sqrt{2} \lambda_{l} a_{lm}(0) /  4\pi$. Inserting this expression into Eq.~\eqref{eq:def density} we obtain
\begin{equation}
      \hat{\rho}(t) = \sum_{lm} \sum_{l^{\prime} m^{\prime}} c_{lm} c_{l^{\prime} m^{\prime}}^{*} \lvert u_{lm} (t )\rangle_{A} \lvert v_{lm} (t )\rangle_{B} \langle u_{l^{\prime} m^{\prime}} (t )\rvert_{A} \langle v_{l^{\prime} m^{\prime}} (t )\rvert_{B}
\end{equation}

\begin{figure}[t]
    \centering
    \captionsetup{
  format=plain,
  margin=1em,
  justification=raggedright,
  }
    \begin{subfigure}{ 0.42\textwidth }
        \centering
        \includegraphics[width = \textwidth]{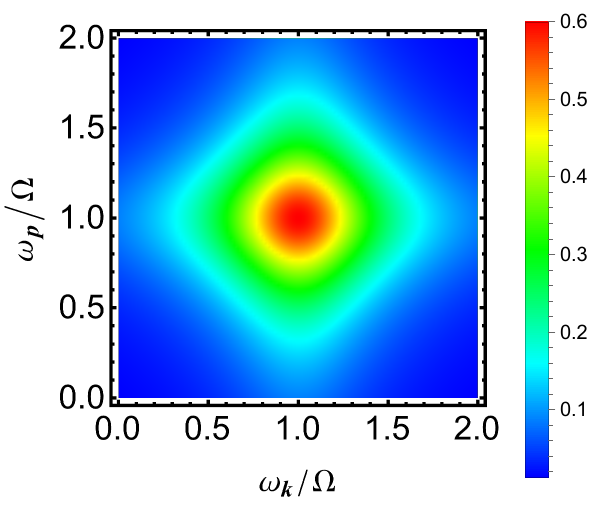}
        \caption{}
        \label{fig:large system two-photon spectrum}
    \end{subfigure}
    \hfill
    \begin{subfigure}{0.52 \textwidth}
        \centering
        \includegraphics[width=\textwidth]{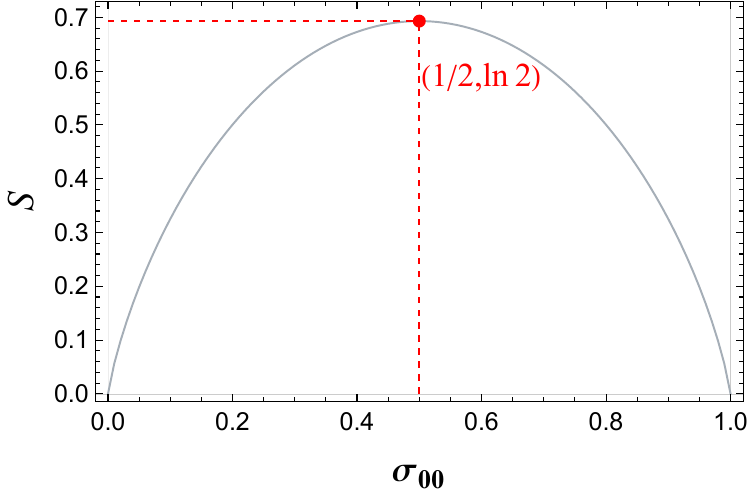}
        \caption{}
        \label{fig:entropy plot}
    \end{subfigure}
  \caption{(a)Two-photon spectrum for a system with the same parameters as in Fig. \ref{fig:large system all probability plot}. (b) von Neumann entropy of the two-photon state containing $s$- and $p_{z}$-wave modes.}
\end{figure}

The reduced density matrix $\hat{\rho}_{A}$ is obtained by performing the trace of $\hat{\rho}(t)$ over the $B$ photon's Hilbert space and using the orthogonality relations Eqs.~\eqref{eq:orthonormal}. We thus obtain
\begin{equation}
    \hat{\rho}_{A}(t) = \operatorname{Tr}_{B} \hat{\rho}(t) =  \sum_{lm} \sigma_{lm} \lvert u_{lm}(t) \rangle \langle u_{lm}(t) \rvert, 
\end{equation}
where $\sigma_{lm} = |c_{lm}|^2$. We note that $\hat{\rho}_{A}(t)$ is diagonal in the $ \lvert u_{lm}\rangle$ basis. The entanglement entropy of the two-photon state is defined as the von Neumann entropy of $\hat{\rho}_{A}(t)$:
\begin{equation}
    S = - \operatorname{Tr} ( \hat{\rho}_{A}(t) \ln \hat{\rho}_{A}(t))=  - \sum_{l, m } \sigma_{lm} \ln \sigma_{lm} . 
    \label{eq: von entropy}
\end{equation}
Evidently, the entropy is nonnegative and vanishes if the state consists of a single mode. Otherwise, the two-photon state is entangled and is maximally entangled when all modes 
are equally probable. We note that the emitted photon pair is entangled only if the initial state is entangled. As an illustrative example, consider a system with only two modes: the $s$ wave and the $p_{z}$ wave. In this case, the entanglement entropy reaches its maximum value when $\sigma_{00} = \sigma_{10} = 1/2$, as shown in Fig.~\ref{fig:entropy plot}.

\subsection{Radiated Power}
\label{sec:radiated power for large system}
\begin{figure}[t]
    \centering
    \captionsetup{
  format=plain,
  margin=1em,
  justification=raggedright,
  }
    \includegraphics[width = 0.75\textwidth]{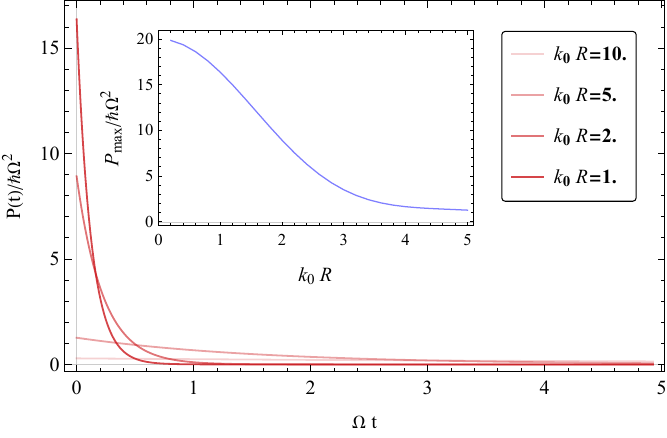}
    \caption{Time-dependence of the radiated power of $s-$wave scattering for different radii. Here $N=100$ and $\Gamma / \Omega =0.1$. The inset shows the maximum value of the power as a function of $k_{0}R$.}
    \label{fig:large system time dependence and scaling with system radius}
\end{figure}

We now compute the  radiated power $P(t)$ which is given by 
\begin{align}
\nonumber
    P(t) &= \hbar \Omega \frac{d}{dt} |b(t)|^2 + 2\hbar \Omega \frac{d}{dt} |c(t)|^2 \\
    &= \frac{\hbar \Omega \Gamma}{ 4 \pi^{2}} \sum\limits_{l,m} \lambda_{l}^{3} \left\vert a_{lm}(0) \right\vert^{2} e^{- \lambda_{l} \Gamma t }.
\end{align}
The time dependence of the power for $s$-wave scattering for various radii $R$ is shown in Fig. \ref{fig:large system time dependence and scaling with system radius}. We see that the power decays monotonically at the rate $\lambda_{l} \Gamma$ for each mode. The maximum value of the power occurs at $t=0$ and is given by
\begin{equation}
    P_{\text{max}} = \frac{\hbar \Omega \Gamma}{ 4 \pi^{2}} \sum\limits_{l,m} \lambda_{l}^{3} \left\vert a_{lm}(0) \right\vert^{2} .
    \label{eq:maximum power for large system}
\end{equation}
As a consistency check, we note that in the limit $k_{0} R \to 0$ in Eq.\eqref{eq:maximum power for large system}, we  recover Eq. \eqref{eq:maximum power for N atom case} for large $N$, as expected. The inset in Fig. \ref{fig:large system time dependence and scaling with system radius} illustrates the relation between the maximum  power $P_{\text{max}}$ and the radius $R$. 

\section{\label{sec:discussion}Discussion}
We have investigated the problem of two-photon collective emission in discrete and continuous atomic systems. Throughout this work, we have employed the rotating wave and pole approximations, the latter being equivalent to the Wigner-Weisskopf approximation. In the discrete case, we discussed the phenomena of stimulated emission and the formation of superradiant and subradiant states for one- and two-atoms, respectively. The two-photon spectrum for the two-atom case revealed strong photon-photon correlations. In the continuous case, we considered a collection of atoms of uniform density in a spherical volume. For small spheres, we found that the decay rate and maximum intensity of the radiated field scaled as the number of atoms $N$. In addition, 
the two-photon spectrum showed strong photon-photon correlations. For large spheres,
the maximum radiated power decreases with system size.  
 
We close by indicating several directions for future research. First, it would be of interest to consider the effects of non-rotating terms in the Hamiltonian. Much is known about this topic for single photon superradiance, where the role played by virtual photons has been emphasized~\cite{Friedberg_Manassah,scully2015single,mirza2016fano}. Virtual transitions are not energy conserving, and can transfer excitations to slowly decaying trapped states and create collective Lamb shifts. A second topic to explore is directional effects in photon emission, for instance in cylindrical volumes. This problem has been studied in the single-photon regime~\cite{cai2016symmetry}, where significant modifications to collective decay rates and  frequency shifts have been found. Finally, in this work we have employed the scalar theory of the electromagnetic field. It would be of interest to generalize our results to setting of the full vector theory. Modifications to the rates of superradiant and subradiant decay in the near-field may be anticipated.


\section*{ACKNOWLEDGEMENTS}
IMM would like to acknowledge financial support from the National Science Foundation grant LEAPS-MPS 2212860, the Miami University College of Arts and Science and Physics Department start-up funding. JCS was supported  by the National Science Foundation grant DMS-1912821 and the AFOSR grant FA9550-19-1-0320.


\appendix
\section{Commutation Relations for Atomic Operators}
\label{appd:Derivation of commutation relation for atomic operators}
The raising and lowering atomic operators for a system of $N$ atoms is defined as the tensor product 
\begin{subequations}
    \begin{align}
    &\hat{\sigma}_{j}  = \openone_{1} \otimes \openone_{2} \otimes \cdots \otimes \ket{0_{j}} \bra{1_{j}} \otimes \cdots \otimes  \openone_{N},\\
    &\hat{\sigma}_{j}^{\dagger} = \openone_{1} \otimes \openone_{2} \otimes \cdots \otimes \ket{1_{j}} \bra{0_{j}} \otimes \cdots \otimes  \openone_{N} ,
    \end{align}
\end{subequations}
where $j=1,\ldots,N$.
Here $\lvert 0_{j} \rangle \left( \lvert 1_{j} \rangle \right)$ denotes the ground (excited) state of the $j$th atom and $\openone_{j}$ is the corresponding identity operator. It is easily seen that $\hat{\sigma}_{i}$ obeys the following anticommutation and commutation relations
\begin{equation}
    \left\{ \hat{\sigma}_{i},\hat{\sigma}_{i}^{\dagger} \right\} = \lvert 0_{i} \rangle \langle 1_{i} \rvert \lvert 1_{i} \rangle \langle 0_{i} \rvert + \lvert 1_{i} \rangle \langle 0_{i} \rvert \lvert 0_{i} \rangle \langle 1_{i} \rvert = \lvert 0_{i} \rangle \langle 0_{i} \rvert + \lvert 1_{i} \rangle \langle 1_{i} \rvert = \openone_i 
\end{equation}
and
\begin{equation}
    \left[ \hat{\sigma}_{i},\hat{\sigma}_{j}^{\dagger} \right] =  \lvert 0_{i} \rangle \langle 1_{i} \rvert \lvert 1_{j} \rangle \langle 0_{j} \rvert - \lvert 1_{j} \rangle \langle 0_{j} \rvert \lvert 0_{i} \rangle \langle 1_{i} \rvert = \lvert 0_{i} 0_{j} \rangle \langle 1_{i} 1_{j} \rvert - \lvert 0_{j} 0_{i} \rangle \langle 1_{j} 1_{i} \rvert = 0 , \quad i\neq j .
\end{equation} 
Here the bosonic nature of the atoms is taken into account so that $\lvert 0_{i} 0_{j}\rangle$ and $\lvert 0_{j} 0_{i}\rangle$ are identified as the same state. We also find that
\begin{equation}
\left\{\hat{\sigma}_{i},\hat{\sigma}_{j} \right\} = 0 , \quad  \left[\hat{\sigma}_{i},\hat{\sigma}_{j} \right] = 0 .
\end{equation}

\vspace{4mm}

\section{Derivation of Equations of Motion}
\label{appd:Equation of Motion}
Here we derive Eq.~\eqref{eq:equation of motion}.
To simplify the calculations, we divide $H \ket{\psi(t)}$ into three terms and calculate each term separately. The first term corresponds to the Hamiltonian acting on the atomic states: 
\begin{equation}
\begin{split}
    &\hat{H} \sum\limits_{ij} a_{ij}(t) \hat{\sigma}_{i} \hat{\sigma}_{j} \lvert 0 \rangle = \left( \sum\limits_{ijl} \hbar \Omega a_{ij}(t) \hat{\sigma}_{l}^\dagger \hat{\sigma}_{l} \hat{\sigma}_i^\dagger \hat{\sigma}_j^\dagger + \sum\limits_{ijl} \sum\limits_{\mathbf{k}} \hbar g a_{ij}(t) e^{- i \mathbf{k} \cdot \mathbf{r}_l } \hat{\sigma}_{l} \hat{a}_{\mathbf{k}}^\dagger \hat{\sigma}_i^\dagger \hat{\sigma}_j^\dagger \right) \lvert 0 \rangle \\
     &= \sum\limits_{ij} 2 \hbar \Omega a_{ij}(t) \hat{\sigma}_{i} \hat{\sigma}_{j} \lvert 0 \rangle  + \sum\limits_{ij} \sum\limits_{\mathbf{k}} \hbar g  a_{ij}(t) \left(e^{-i \mathbf{k} \cdot \mathbf{r}_i} \hat{\sigma}_{j} \hat{\mathbf{a}}_{\mathbf{k}} \lvert 0 \rangle+e^{-i \mathbf{k} \cdot \mathbf{r}_j} \hat{\sigma}_{i} \hat{a}_{\mathbf{k}} \lvert 0 \rangle \right).
\end{split} 
\end{equation}
The second term accounts for the action of Hamiltonian on the mixed-state amplitude:
\begin{equation}
    \begin{split}
    &\hat{H} \sum\limits_{i \mathbf{k}} b_{i \mathbf{k}}(t) \hat{\sigma}_{i} \hat{a}_{\mathbf{k}} \lvert 0 \rangle = \Bigg[ \sum\limits_l \hbar \Omega \hat{\sigma}_{l}^\dagger \hat{\sigma}_{l} + \sum\limits_{\mathbf{p}} \hbar \omega_{\mathbf{p}} \hat{a}_{\mathbf{p}}^\dagger \hat{a}_{\mathbf{p}} + \hbar g \sum\limits_j \sum\limits_{\mathbf{p}} \left(e^{i \mathbf{p}\cdot \mathbf{r}_j } \hat{\sigma}_j^\dagger \hat{a}_{\mathbf{p}} \right. \left. + e^{-i \mathbf{k} \cdot \mathbf{r}_j } \hat{\sigma}_j \hat{a}_{\mathbf{k}}^\dagger \right) \Bigg] \\
    &\times\sum\limits_{i,\mathbf{k}} b_{i, \mathbf{k}}(t) \hat{\sigma}_i^\dagger \hat{a}_{\mathbf{k}}^\dagger \lvert 0 \rangle = \sum\limits_{i,\mathbf{k}} \hbar \left(\Omega+\omega_{\mathbf{k}}\right) b_{i\mathbf{k}}(t) \hat{\sigma}_{i} \hat{a}_{\mathbf{k}} \lvert 0 \rangle+ \hbar g \sum\limits_{ij} \sum\limits_{\mathbf{k}} (1-\delta_{ij})b_{i\mathbf{k}}(t) e^{i \mathbf{k} \cdot \mathbf{r}_j} \hat{\sigma}_{i} \hat{\sigma}_{j} \lvert 0 \rangle\\
    &\quad\quad+ \hbar g \sum\limits_i \sum\limits_{\mathbf{k} \mathbf{p}} b_{i \mathbf{k}}(t) e^{-i \mathbf{p}\cdot \mathbf{r}_i} \hat{a}_{\mathbf{k}} \hat{a}_{\mathbf{p}} \lvert 0 \rangle.
    \end{split}
\end{equation}
Finally, the third term corresponds to the  Hamiltonian acting on the two-photon state: 
\begin{equation}
    \begin{split}
    &\hat{H} \sum\limits_{\mathbf{k} \mathbf{p}} c_{\mathbf{k},\mathbf{p}}(t) \hat{a}_{\mathbf{k}} \hat{a}_{\mathbf{p}} \lvert 0 \rangle = \left( \sum\limits_{\mathbf{q}} \hbar \omega_{\mathbf{q}} \hat{a}_{\mathbf{q}}^{\dagger} \hat{a}_{\mathbf{q}} +\hbar g \sum\limits_{i} \sum\limits_{\mathbf{q}} e^{i \mathbf{q} \cdot \mathbf{r}_i} \hat{\sigma}_i^\dagger \hat{a}_{\mathbf{q}} \right) \sum\limits_{\mathbf{k},\mathbf{p}} c_{\mathbf{k},\mathbf{p}}(t) \hat{a}^{\dagger}_{\mathbf{k}}\hat{a}^{\dagger}_{\mathbf{p}} \lvert 0 \rangle\\
    & = \sum\limits_{\mathbf{k},\mathbf{p}} \hbar \left(\omega_{\mathbf{k}}+\omega_{\mathbf{p}}\right) c_{\mathbf{k} \mathbf{p}}(t) \hat{a}_{\mathbf{k}} \hat{a}_{\mathbf{p}} \lvert 0 \rangle + \hbar g \sum\limits_{i} \sum\limits_{\mathbf{k},\mathbf{p}} c_{\mathbf{k},\mathbf{p}}(t) \left(e^{i \mathbf{k} \cdot \mathbf{r}_i} \hat{\sigma}_{i} \hat{a}_{\mathbf{p}} \lvert 0 \rangle+ e^{i \mathbf{p}\cdot \mathbf{r}_i} \hat{\sigma}_{i} \hat{a}_{\mathbf{k}} \lvert 0 \rangle \right).
    \end{split}
\end{equation}
Putting everything together, we find
\begin{equation}
    \begin{split}
    &\hat{H} \lvert  \psi(t) \rangle\\
    &= \sum\limits_{ij} \left(2\hbar \Omega a_{ij}(t) + \hbar g \sum\limits_{\mathbf{k}} (1-\delta_{ij})b_{i \mathbf{k}}(t) e^{i \mathbf{k} \cdot \mathbf{r}_j}\right) \hat{\sigma}_{i} \hat{\sigma}_{j} \lvert 0 \rangle\\
    &+ \sum\limits_{\mathbf{k},\mathbf{p}} \left[\hbar \left(\omega_{\mathbf{k}}+\omega_{\mathbf{p}}\right) c_{\mathbf{k} \mathbf{p}}(t)+\hbar g \sum\limits_i b_{i\mathbf{k}}(t) e^{-i \mathbf{p}\cdot \mathbf{r}_i}\right] \hat{a}_{\mathbf{k}} \hat{a}_{\mathbf{p}} \lvert 0 \rangle + \sum\limits_{i,\mathbf{k}} \left[\hbar \left(\Omega+\omega_{\mathbf{k}} \right) b_{i \mathbf{k}}(t) \right.\\
    & \left. +\hbar g \sum\limits_j \left(a_{ij}(t)+a_{ji}(t)\right) e^{-i \mathbf{k} \cdot \mathbf{r}_j} + \hbar g \sum\limits_{\mathbf{p}} \left(c_{\mathbf{k} \mathbf{p}}(t)+c_{\mathbf{p}\mathbf{k}}(t)\right) e^{i \mathbf{p}\cdot \mathbf{r}_i}\right] \hat{\sigma}_{i} \hat{a}_{\mathbf{k}} \lvert 0 \rangle.
    \end{split}
\end{equation}
Finally, by using the symmetry of $a_{ij}(t)$ and $c_{\mathbf{k}\mathbf{p}}(t)$ and projecting from the left-hand side  with  $ \hat{\sigma}_{i} \hat{\sigma}_{j} \lvert 0 \rangle, ~\hat{\sigma}_{i} \hat{a}_{\mathbf{k}} \lvert 0 \rangle, ~\text{and}~ \hat{a}_{\mathbf{k}} \hat{a}_{\mathbf{p}} \lvert 0 \rangle  $, we obtain Eq.~\eqref{eq:equation of motion}.


\section{\label{appd:calculation of self-energy}Calculation of the Self-energy }
Here we calculate the self-energy
\begin{equation}
\Sigma = g^2 \sum\limits_{\mathbf{p}}\frac{1}{\Omega-\omega_{\mathbf{p}}} . 
\end{equation}
To proceed, we make use of the identity
\begin{equation}
\lim\limits_{\epsilon \rightarrow 0^{+}} \frac{1}{x+i \epsilon}=P\frac{1}{x}-i \pi \delta(x) ,
\end{equation}
where  $P$ denotes the principal value.
We then obtain
\begin{equation}
\Sigma = g^2 \left(P \sum\limits_{\mathbf{k}}\frac{1}{\Omega-\omega_{\mathbf{k}}} - i \pi \sum\limits_{\mathbf{k}} \delta(\Omega-\omega_{\mathbf{k}})\right) .
\end{equation}
The quantity ${\Im}~\Sigma$ is given by
\begin{eqnarray}
     \Im \Sigma &=& -g^2 \pi \sum\limits_{\mathbf{k}} \delta(\Omega-\omega_{\mathbf{k}}) = -g^2 \pi \frac{V}{(2 \pi)^3} \int d^3{k} \delta(\Omega-\omega_{\mathbf{k}})\nonumber\\
    &=& -\frac{g^2 V \Omega^2}{2 \pi c^3}. 
\end{eqnarray}
Likewise ${\Re}~\Sigma$ is given by
\begin{eqnarray}
\nonumber
{\Re}~\Sigma &=& g^2 P \sum\limits_{\mathbf{k}}\frac{1}{\Omega-\omega_{\mathbf{k}}} =  g^2 \frac{V}{(2 \pi)^3} P \int d^3 {k} \frac{1}{\Omega-\omega_{\mathbf{k}}} \\
    &=& g^2 \frac{V}{(2 \pi)^3} 4 \pi P\int_0^{\frac{2\pi}{\Lambda}} dk \frac{k^2}{k_0-k} \; \nonumber\\
    &=& -\frac{g^2 V}{2 \pi^2 c} \left( \frac{2 \pi^2}{\Lambda^2} + \frac{2\pi}{\Lambda} k_0 - \ln\left( \frac{k_0}{\frac{2\pi}{\Lambda}-k_0}\right) \right) ,
\end{eqnarray}
where we have introduced a cutoff ${2\pi}/{\Lambda}$ to regularize the divergence and $k_0 = {\Omega}/{c} < {2\pi}/{\Lambda}$. We summarize the above as
\begin{equation}
    \Sigma = \delta \omega - i \Gamma/2,
\end{equation}
where
\begin{subequations}
    \begin{align}
        \delta \omega =&-\frac{g^2 V}{2 \pi^2 c} \left[ \frac{2 \pi^2}{\Lambda^2} + \frac{2\pi}{\Lambda} k_0 - \ln\left( \frac{k_0}{\frac{2\pi}{\Lambda}-k_0}\right) \right], \\
        \Gamma =& \frac{g^2 V \Omega^2}{ \pi c^3} .
    \end{align}
\end{subequations}
The quantities $\delta \omega$ and $\Gamma$ are the Lamb shift and the 
atomic decay rate, respectively.
\vspace{4mm}

\section{\label{appd:calculation of interaction energy}Calculation of the Interaction Energy}
In this section, we will calculate the interaction energy, which in the pole approximation
is of the form
\begin{equation}
    \Delta_{jl} = g^2 \sum\limits_{\mathbf{k}}  \frac{e^{i \mathbf{k} \cdot \mathbf{r}_{jl}}}{\Omega-\omega_{\mathbf{k}}+ i\epsilon},
\end{equation}
where $\mathbf{r}_{jl} = \mathbf{r}_j - \mathbf{r}_l$. Evidently $\Delta_{jl} = \Delta_{lj}$.
Making use of the identity
\begin{equation}
\lim\limits_{\epsilon \rightarrow 0^{+}} \frac{1}{x+i \epsilon}=P\frac{1}{x}-i \pi \delta(x) ,
\end{equation}
we obtain
\begin{equation}
    \Delta_{jl} = g^2 \left( P \sum\limits_{\mathbf{k}} \frac{e^{i \mathbf{k} \cdot \mathbf{r}_{jl}}}{\Omega-\omega_{\mathbf{k}}} -i\pi \sum\limits_{\mathbf{k}} e^{i \mathbf{k} \cdot \mathbf{r}_{jl}} \delta(\Omega-\omega_{\mathbf{k}}) \right) .
\end{equation}
It follows that
\begin{eqnarray}
    {\Im} \Delta_{jl} &=& -\pi \sum\limits_{\mathbf{k}} e^{i \mathbf{k} \cdot \mathbf{r}_{jl}} \delta(\Omega-\omega_{\mathbf{k}}) = -g^2 \pi \frac{V}{(2 \pi)^3} \int d^3 {k} e^{i \mathbf{k} \cdot \mathbf{r}_{jl}} \delta(\Omega-\omega_{\mathbf{k}})\nonumber\\
    &=& -\frac{\Gamma}{2}\text{sinc}(k_0 r_{jl}) , 
\end{eqnarray}
where $k_0=\Omega/c$ and $\Gamma = {g^2 V \Omega^2}/{ \pi c^3}$. 
We also have
\begin{eqnarray}
    {\Re} \Delta_{jl} &=& g^2 P \sum\limits_{\mathbf{k}} \frac{e^{i \mathbf{k} \cdot \mathbf{r}_{jl}}}{\Omega-\omega_{\mathbf{k}}} = g^2  \frac{V}{(2 \pi)^3} P\int d^3 {k} \frac{e^{i \mathbf{k} \cdot \mathbf{r}_{jl}}}{\Omega-\omega_{\mathbf{k}}}\nonumber\\
     &=&  \frac{g^2V}{2 \pi^2 c r_{jl}} P\int_0^{\frac{2\pi}{\Lambda}} dk \; \frac{k \sin (k r_{jl})}{k_0 - k} , 
\end{eqnarray}
where we have introduced a cutoff to regularize the divergence. If $k_0 r_{jl} \ll 1$ then
\begin{eqnarray*}
     {\Re} \Delta_{jl} \simeq \frac{\Gamma}{2 \pi} \left[ \frac{\cos\left({\frac{2\pi r_{jl}}{\Lambda}}\right)-1}{k_0^2 r_{jl}^2} + \frac{\pi}{2 k_0 r_{jl}} + \ln({k_0 r_{jl}})\right].
\end{eqnarray*}
Putting everything together we find that
\begin{eqnarray}
    \Delta_{jl} 
    &=& \delta \omega_{jl} - i \frac{\Gamma_{jl}}{2} ,
\end{eqnarray}
where 
\begin{eqnarray}
     \delta \omega_{jl} &=& \frac{\Gamma}{2 \pi} \left[ \frac{\cos{\frac{2\pi R_{jl}}{\Lambda}}-1}{k_0^2 R_{jl}^2} + \frac{\pi}{2 k_0 R_{jl}} + \ln{\left(k_0 R_{jl}\right)}\right] ,\nonumber \\ 
     \Gamma_{jl} &=& \Gamma \text{sinc} \left(k_0 R_{jl}\right) .
\end{eqnarray}

\section{\label{appd:two atom problem}Two Atom Case}
In this section, we derive the mode-independent probabilities presented in Eq.~\eqref{twoatomres}. The probability of the two-atom state can be calculated directly as follows:
\begin{equation}
    \left\vert a(t) \right\vert^{2} = 2( \left\vert a_{12}(t)  \right\vert^{2} + \left\vert a_{21}(t)  \right\vert^{2})  = e^{-2\Gamma t}.
\end{equation}
We then compute the probability of the mixed state
\begin{align}
    \left\vert b(t) \right\vert^{2} &= \sum\limits_{\mathbf{k}}\left( \left\vert b_{1\mathbf{k}}(t)\right\vert^{2} + \left\vert  b_{2\mathbf{k}}(t) \right\vert^{2} \right)\nonumber\\
    &= \frac{1}{2\pi} \int_{-\infty}^{\infty} d\omega_{\mathbf{k}} \left[ \frac{\Gamma_{+}\left( e^{-2\Gamma t}-2\cos(\Omega-\omega_{\mathbf{k}})t e^{-\Gamma t-\Gamma_{+}t/2}+e^{-\Gamma_{+}t} \right)}{(\Omega-\omega_{\mathbf{k}})^{2}+\Gamma_{-}^{2} / 4} +(\Gamma_{+} \leftrightarrow \Gamma_{-})\right] \nonumber \\
    &  = \frac{\Gamma_{+}}{\Gamma_{-}} (e^{-\Gamma_{+}t} - e^{-2\Gamma t})+\frac{\Gamma_{-}}{\Gamma_{+}} (e^{-\Gamma_{-}t} - e^{-2\Gamma t}).
\end{align}
Lastly, the probability of the two-photon state is given by
\begin{align}
    \left\vert c(t) \right\vert^{2} =& 2\sum\limits_{\mathbf{k},\mathbf{p}} \left\vert c_{\mathbf{k}\mathbf{p}}(t) \right\vert^{2}\nonumber\\
    =& \frac{1}{8\pi^{2}} \iint d\omega_{\mathbf{k}} d\omega_{\mathbf{p}} \Bigg\lbrace \Gamma_{+}^{2} \Bigg| \frac{e^{i(2\Omega-\omega_{\mathbf{k}}-\omega_{\mathbf{p}})t}}{2\Omega-\omega_{\mathbf{k}}-\omega_{\mathbf{p}}-i \Gamma}\left( \frac{1}{\Omega-\omega_{\mathbf{k}}-i\Gamma_{+} /2}+\frac{1}{\Omega-\omega_{\mathbf{p}}-i\Gamma_{+} /2} \right)\nonumber\\
        &- \frac{e^{i(\Omega-\omega_{\mathbf{k}})t-\Gamma_{+}t / 2 }}{\left( \Omega-\omega_{\mathbf{k}}-i \Gamma_{-} / 2 \right)\left( \Omega-\omega_{\mathbf{p}}-i \Gamma_{+} / 2 \right)}-\frac{e^{i(\Omega-\omega_{\mathbf{p}})t-\Gamma_{+}t / 2}}{\left( \Omega-\omega_{\mathbf{k}}-i \Gamma_{+} / 2 \right)\left( \Omega-\omega_{\mathbf{p}}-i \Gamma_{-} / 2 \right)}\nonumber \\
        &\left. + \frac{e^{-\Gamma t}}{2\Omega-\omega_{\mathbf{k}}-\omega_{\mathbf{p}}-i\Gamma} \left( \frac{1}{\Omega-\omega_{\mathbf{k}}-i\Gamma_{-} / 2} + \frac{1}{\Omega-\omega_{\mathbf{p}}-i\Gamma_{-} / 2}\right)\Bigg|^{2}+\left(\Gamma_{+} \leftrightarrow \Gamma_{-} \right) \right\} \nonumber \\
         =& 
        1 - \left(1 - \frac{\Gamma_{+}^{2} + \Gamma_{-}^{2}}{\Gamma_{+}\Gamma_{-}}\right) e^{-2\Gamma t } - \frac{\Gamma_{+}}{\Gamma_{-}} e^{-\Gamma_{+}t} -\frac{\Gamma_{-}}{\Gamma_{+}} e^{-\Gamma_{-}t}.
\end{align}

\section{Small System \label{appd:small system}}
\subsection{Derivation of the Atomic Amplitude}
Here we derive Eq.~\eqref{eq:solution to N atom for atomic amplitude}. 
We begin by recalling Eq.~\eqref{eq:aij}:
\begin{align}
(is - 2\Omega)a_{ij}(s) = i a_{ij}(0) +  (1-\delta_{ij}) \Sigma  \sum_{l}\left(a_{il}(s) + a_{jl}(s) \right).
\label{eq:atomic amplitude equation of motion aij}
\end{align}
By summing over $j$, we obtain
\begin{align}
(is - 2\Omega)\sum_{j}a_{ij}(s) = i \sum_{j} a_{ij}(0) +  (N-2) \Sigma  \sum_{l} a_{il}(s)  +  \Sigma \sum_{j} \sum_{l} a_{jl}(s).
\label{eq:sum over j}
\end{align}
Then, we proceed by summing over $i$ to obtain
\begin{equation}
    (is - 2\Omega) \sum_{i} \sum_{j} a_{ij}(s) = i \sum_{i} \sum_{j} a_{ij}(0) +  (N-2) \Sigma \sum_{i} \sum_{l} a_{il}(s)  +  N \Sigma \sum_{j} \sum_{l} a_{jl}(s) .
    \label{eq:sum over i and j}
\end{equation}
Taking into account the initial conditions $a_{ij}(0) =1  / \sqrt{2 N (N-1) }$ for $i\neq j$ and $a_{ii} (0) = 0$, we find that
\begin{equation}
    \sum_{i,j}  a_{ij}(s) = \sqrt{ \frac{N(N-1)}{2}} \frac{ i }{ is - 2\Omega - 2(N-1) \Sigma}.
    \label{eq:sum aij over i j expression}
\end{equation}
Substituting this result into Eq.~\eqref{eq:sum over j} yields
\begin{equation}
    \sum_{j} a_{ij}(s) = \sqrt{\frac{N-1}{2 N}} \frac{i}{ i s - 2 \Omega - 2 (N-1) \Sigma},
    \label{eq:sum aij over j expression}
\end{equation}
for all $i$. Finally, substituting Eq. \eqref{eq:sum aij over j expression} into Eq. \eqref{eq:atomic amplitude equation of motion aij}, we arrive at
\begin{equation}
    a_{ij}(s) = \frac{1}{\sqrt{2 N (N-1)}} \frac{i}{ is - 2 \Omega - 2 (N-1) \Sigma} ,~(i \neq j)
\end{equation}
which is Eq.~\eqref{eq:solution to N atom for atomic amplitude} in the main text.

\subsection{Computation of Mode-Independent Probabilities}
Here we derive the mode-independent probabilities presented in Eq.~ \eqref{eq:N_atom_probabilities}. The probability of the two-atom state can be calculated directly as follows:
\begin{equation}
    \left\vert a(t) \right\vert^{2} = 2 \sum\limits_{i \neq j} \left\vert a_{ij}(t) \right\vert^{2} = e^{-2(N-1)\Gamma t }.
\end{equation}
Next, we compute the probability of the mixed state, which is given by
\begin{align}
    \left\vert b(t) \right\vert^{2} &= \sum\limits_{i , \mathbf{k}} \left\vert b_{i \mathbf{k}}(t) \right\vert^{2} \nonumber\\
     &= (N-1) \frac{\Gamma}{\pi} \int_{-\infty}^{\infty} d \omega_{\mathbf{k}} \frac{e^{-2(N-1)\Gamma t }-2 \cos(\Omega-\omega_{\mathbf{k}})t e^{-(\frac{3}{2}N-1)\Gamma t }+e^{-N \Gamma t }}{\left( \Omega-\omega_{\mathbf{k}} \right)^{2}+(N-2)^{2} \Gamma^{2} /4} \nonumber \\
    &= 2\frac{N-1}{N-2} \left[ e^{-N \Gamma t } - e^{- 2(N-1) \Gamma t } \right].
\end{align}
Finally, we calculate the probability of the two-photon state as follows:
\begin{align}
    &\left\vert c(t) \right\vert^{2} =2\sum\limits_{\mathbf{k},\mathbf{p}} \left\vert c_{\mathbf{k}\mathbf{p}}(t) \right\vert^{2} = \frac{N(N-1)\Gamma^{2}}{4\pi^{2}} \int_{-\infty}^{\infty} d \omega_{\mathbf{k}} \int_{-\infty}^{\infty} d \omega_{\mathbf{p}}\times \nonumber\\
    & \left| \frac{e^{i(2\Omega-\omega_{\mathbf{k}}-\omega_{\mathbf{p}})t}}{2\Omega-\omega_{\mathbf{k}}-\omega_{\mathbf{p}}-i(N-1)\Gamma} \left( \frac{1}{\Omega-\omega_{\mathbf{k}}-i N \Gamma /2}+\frac{1}{\Omega-\omega_{\mathbf{p}}-i N \Gamma /2} \right)\right.\nonumber\\
    &+ \frac{e^{-(N-1)\Gamma t }}{2\Omega-\omega_{\mathbf{k}}-\omega_{\mathbf{p}}-i(N-1)\Gamma} \left( \frac{1}{\Omega-\omega_{\mathbf{k}}-i (N-2) \Gamma /2}+\frac{1}{\Omega-\omega_{\mathbf{p}}-i (N-2) \Gamma /2} \right)\nonumber\\
    &\left. - \frac{e^{i(\Omega-\omega_{\mathbf{k}})t- \frac{1}{2} N \Gamma t }}{\left( \Omega-\omega_{\mathbf{p}}-iN \frac{\Gamma}{2} \right)\left( \Omega-\omega_{\mathbf{k}}-i(N-2) \frac{\Gamma}{2} \right)}- \frac{e^{i(\Omega-\omega_{\mathbf{p}})t- \frac{1}{2} N \Gamma t }}{\left( \Omega-\omega_{\mathbf{k}}-iN \frac{\Gamma}{2} \right)\left( \Omega-\omega_{\mathbf{p}}-i(N-2) \frac{\Gamma}{2} \right)} \right|^{2} \nonumber\\
     =& 1- 2 \frac{N-1}{N-2} e^{-N \Gamma t } + \frac{N}{N-2} e^{-2(N-1) \Gamma t } .
\end{align}

\section{\label{appd:continuum}Large System}

\subsection{Derivation of the Atomic Amplitude}
Here we derive Eq. \eqref{eq:large system solution for a}.
We begin by recalling Eq. \eqref{eq:equation for a in lagre system}:
\begin{equation}
   (is - 2\Omega) a(\mathbf{x},\mathbf{y},s) = i a(\mathbf{x},\mathbf{y},0) + \rho_{0} \int_{V_{a}} d^{3} {z} \left[ \Delta(\mathbf{y},\mathbf{z}) a(\mathbf{z},\mathbf{x},s) +  (\mathbf{x} \leftrightarrow \mathbf{y}) \right].
\end{equation}
Substituting the expression for $a(\mathbf{x},\mathbf{y},s)$ given by Eq. \eqref{eq:expansion of a} into the above, we obtain
\begin{equation}
    \begin{aligned}
    &\sum\limits_{l,m} \left[ (is - 2\Omega) a_{lm}(s) - i a_{lm}(0) \right] j_{l}(k_{0}x)  j_{l}(k_{0}y) Y_{lm}(\mathbf{\hat{x}}) Y_{lm}^{*}(\mathbf{\hat{y}})  \\
    &= -2 \pi i \Gamma \rho_{0}  \int_{V_{a}} d^{3} \mathbf{z} \sum\limits_{l,m} j_{l}(k_{0}y) j_{l}(k_{0}z) Y_{lm}(\mathbf{\hat{y}})Y_{lm}^{*}(\mathbf{\hat{z}})\\
    &\times \sum\limits_{l'm'} a_{l' m'}(s) j_{l'}(k_{0}y)  j_{l'}(k_{0}z) Y_{l'm'}(\mathbf{\hat{z}}) Y_{l'm'}^{*}(\mathbf{\hat{x}}) + (\mathbf{x} \leftrightarrow \mathbf{y}). \\
    \end{aligned}
\end{equation}
Next, we make use of  the orthogonality of the spherical harmonics,
\begin{equation}
    \int d \mathbf{\hat{z}}~ Y_{lm}^{*}(\mathbf{\hat{z}}) Y_{l'm'}(\mathbf{\hat{z}}) = \delta_{ll'} \delta_{mm'},
\end{equation}
thereby obtaining
\begin{equation}
    \sum\limits_{l,m} \left[ (is - 2\Omega + i \lambda_{l} \Gamma) a_{lm}(s) - i a_{lm}(0) \right] j_{l}(k_{0}x)  j_{l}(k_{0}y) Y_{lm}(\mathbf{\hat{x}}) Y_{lm}^{*}(\mathbf{\hat{y}}) = 0,
\end{equation}
where
\begin{equation}
    \lambda_{l} = 4 \pi \rho_{0} \int_{0}^{R} d r ~ r^{2} j_{l}^{2}(k_{0}r). 
\end{equation}
It follows that
\begin{equation}
    a_{lm}(s) = \frac{i a_{lm}(0)}{is - 2\Omega + i \lambda_{l} \Gamma}.
\end{equation}
Therefore, $a(\mathbf{x},\mathbf{y},s)$ is given by 
\begin{equation}
    a(\mathbf{x},\mathbf{y},s) =  \sum\limits_{l,m} \frac{i a_{lm}(0)}{is - 2\Omega + i \lambda_{l} \Gamma} j_{l}(k_{0}x)  j_{l}(k_{0}y) Y_{lm}(\mathbf{\hat{x}}) Y_{lm}^{*}(\mathbf{\hat{y}}),
\end{equation}
which is Eq. \eqref{eq:large system solution for a} in the main text.

\subsection{\label{appd:prob large system}Computation of Mode-Independent Probabilities}
Here we derive the mode-independent probabilities presented in Eq.~\eqref{eq:probabilities for large system}. The probability of the two-atom state can be calculated directly as follows:
\begin{equation}
    \left\vert a(t) \right\vert^{2} = 2 \rho_{0}^{2} \int_{V_{a}} d^{3} {x} \int_{V_{a}} d^{3} {y} \left\vert a(\mathbf{x},\mathbf{y},t) \right\vert^{2} = \frac{1}{8\pi^{2}} \sum_{l,m} \lambda_{l}^{2} \left\vert a_{lm}(0) \right\vert^{2}  e^{-2 \lambda_{l} \Gamma t}.
\end{equation}
Next, the probability of the mixed state
\begin{align}
    \left\vert b(t) \right\vert^{2} &=  \frac{\rho_{0}V}{(2\pi)^{3}} \int_{V_{a}} d^{3} {x} \int d^{3} {k} ~ \left\vert b(\mathbf{x},\mathbf{k},t) \right\vert^{2} \nonumber\\
        &= \frac{\Gamma}{ 8 \pi^{3}} \sum_{lm}  \lambda_{l}^{3} \left\vert a_{lm}(0) \right\vert^{2} \int_{-\infty}^{\infty} d \omega_{\mathbf{k}} \frac{ e^{- 2 \lambda_{l} \Gamma t } - 2 \cos(\Omega - \omega_{\mathbf{k}})t e^{- \frac{3}{2} \lambda_{l} \Gamma t } + e^{- \lambda_{l} \Gamma t }}{ (\Omega - \omega_{\mathbf{k}})^{2} +  \lambda_{l}^{2} \Gamma^{2} / 4} \nonumber\\
        &= 
        \frac{1}{4\pi^{2}}
        \sum_{l,m} \lambda_{l}^{2} \left\vert a_{lm}(0) \right\vert^{2} (e^{-\lambda_{l} \Gamma t } - e^{-2 \lambda_{l} \Gamma t }) . 
\end{align}
Finally, we calculate the probability of the two-photon state
\begin{align}
    \left\vert c(t) \right\vert^{2} &= \frac{2 V^{2}}{(2\pi)^{6}} \int d^{3}{k} d^{3} {p} \left\vert c(\mathbf{k},\mathbf{p},t) \right\vert^{2}= \frac{\Gamma^{2}}{32 \pi^{4}} \sum\limits_{l,m} \lambda_{l}^{4} \left\vert a_{lm}(0) \right\vert^{2} \nonumber\\
    &\quad \times \int_{-\infty}^{\infty} d\omega_{\mathbf{k}} \int_{-\infty}^{\infty} d\omega_{\mathbf{p}} \frac{ \left\vert e^{-\lambda_{l} \Gamma t} + e^{i(2\Omega - \omega_{\mathbf{k}} - \omega_{\mathbf{p}})t} - \left( e^{i(\Omega - \omega_{\mathbf{k}})t } + e^{i(\Omega - \omega_{\mathbf{p}})t } \right) e^{- \frac{1}{2} \lambda_{l} \Gamma t } \right\vert^{2} }{\left[ (\Omega - \omega_{\mathbf{k}})^{2} + \lambda_{l}^{2} \Gamma^{2} / 4 \right] \left[ (\Omega - \omega_{\mathbf{p}})^{2} + \lambda_{l}^{2} \Gamma^{2} / 4\right]} \nonumber \\
    & = \frac{1}{8\pi^{2}} \sum\limits_{l,m} \lambda_{l}^{2} \left\vert a_{lm}(0) \right\vert^{2}(1- 2 e^{- \lambda_{l} \Gamma t } + e^{- 2 \lambda_{l} \Gamma t }) .
\end{align}

\bibliographystyle{ieeetr}
\bibliography{paper}
\end{document}